\documentclass{jfm}
\usepackage[utf8]{inputenc}
\usepackage[english]{babel}
\usepackage{graphicx}
\usepackage{adjustbox}
\usepackage{subcaption}
\usepackage{comment}
\usepackage{amsmath}
\usepackage{amssymb}
\usepackage[colorlinks=true,citecolor=blue,linkcolor=blue,urlcolor=blue]{hyperref}
\usepackage{epstopdf, epsfig}
\usepackage{textcomp,gensymb} 
\usepackage{natbib}

\newcommand{\ReNumber}{\mathrm{Re}}
\newcommand{\WeNumber}{\mathrm{We}}
\newcommand{\DR}{\mathrm{DR}}

\shorttitle{Roughness on liquid-infused surfaces due to capillary waves}
\shortauthor{J. Sundin, S. Zaleski and S. Bagheri}

\title{Roughness on liquid-infused surfaces induced by capillary waves}

\author{Johan Sundin\aff{1}\corresp{\email{johasu@mech.kth.se}}, Stéphane Zaleski\aff{2,3} \and Shervin Bagheri\aff{1} }

\affiliation{\aff{1}Linné FLOW Centre, Dept.~Engineering Mechanics, KTH, Stockholm, Sweden \aff{2}Sorbonne Universit\'e \& CNRS, Institut Jean Le Rond d'Alembert, UMR 7190, Paris, France \aff{3}Institut Universitaire de France, Institut Jean Le Rond d’Alembert, UMR 7190, Paris, France}

\begin{document}
\maketitle
\begin{abstract}
Liquid-infused surfaces (LIS) are a promising technique for reducing friction, fouling and icing in both laminar and turbulent flows. Previous work has demonstrated that these surfaces are susceptible to shear-driven drainage. Here, we report a different failure mode using direct numerical simulations of a turbulent channel flow with liquid-infused longitudinal grooves.  When the liquid-liquid surface tension is small and/or grooves are wide, we observe travelling-wave perturbations on the interface with amplitudes  larger than the viscous sublayer of the turbulent flow. These capillary waves induce a roughness effect that increases  drag. The generation mechanism of these waves is explained using the theory developed by Miles for gravity waves. Energy is transferred from the turbulent flow to the LIS provided that there is a negative curvature of the mean flow at the critical layer. Given the groove width, the Weber number and an estimate of the friction Reynolds number, we provide relations to determine whether a LIS behaves as a smooth or rough surface in a turbulent flow.
 \end{abstract}


\vspace{-0.5cm}
\section{Introduction}
A protective and functional surface coating can be created by lubricating a textured surface with an appropriate liquid. The drag-reducing properties of these liquid-infused surfaces (LIS) have been explored recently both numerically \citep{fu17, cartagena18, arenas19} and experimentally \citep{buren17, fu19}.
LIS can also prevent fouling \citep{epstein12}, corrosion \citep{wang15} and ice formation \citep{kim12}.

The drag-reducing capabilities of LIS for liquid flows is often compared to those of superhydrophobic surfaces (SHS), where air is used as the infused medium. The low viscosity of air is beneficial for drag reduction, but the use of SHS in turbulent applications is restricted by mass diffusion \citep{ling17} and instability of the gas pockets \citep{seo18}. 
For LIS, the mass diffusion is negligible if the liquids are immiscible, and LIS are not susceptible to failure due to hydrostatic pressure \citep{wong11}. However, also for LIS, the stability of the interface depends on the texture's geometry, the surface tension between the two liquids and the contact angle at the liquid-liquid-solid interface. In particular, these surfaces may experience shear-driven drainage of the infused liquid, but this can be mitigated, for example with chemical patterning \citep{wexler15b, fu19}. 

In this paper, we show that capillary motion of the liquid-liquid interface may drastically lower the drag-reducing performance of LIS. 
In the present study, 
%
the surface texture is fixed to longitudinal (streamwise-aligned) grooves. We use direct numerical simulations of a LIS in a liquid turbulent channel flow for frictional Reynolds numbers around $\ReNumber_\tau \approx 180$. The employed volume-of-fluid (VOF) framework  allows for large interface deformations (low surface tension) and moving contact lines. %
When the liquid-liquid surface tension is small and/or grooves are wide, we find travelling-wave perturbations on the interface with amplitudes  larger than the viscous sublayer of the turbulent flow ($a^+\approx 5-8$). These capillary waves induce a roughness effect and increase friction drag.

The detrimental capillary waves develop for viscosity and density ratios of one, which excludes interface instability mechanisms driven by density and viscosity stratification \citep{boomkamp96}.
 Instead, we find that 
the linear instability can be described by the theory developed by \citet{miles57} in the context of two-dimensional gravity waves. 
%
This inviscid instability is due to an energy transfer from the external flow to the waves that makes them grow in time at an exponential rate. The energy transfer can occur if (i) there is a negative curvature of the mean velocity profile where it equals the phase speed of a wave, i.e.~at the critical layer and, (ii) the critical layer is not too far away from the surface, so that velocity fluctuations due to the wave (dispersive stresses) are non-zero. 
%

\begin{figure}
  \centering
  \begin{minipage}{0.45\textwidth}
  \centering
  \begin{subfigure}{\textwidth}
     \centering
     \includegraphics[width=4.6cm]{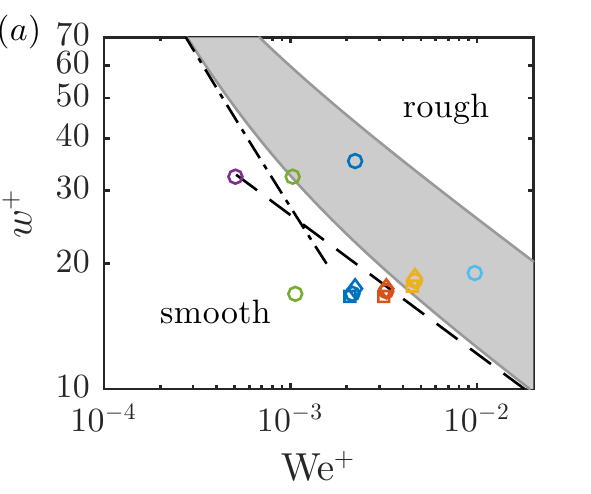}
     \captionlistentry{}
     \label{fig:stabilityMap} 
  \end{subfigure}
  \begin{subfigure}{\textwidth}
      \raggedleft
      \includegraphics[width=4.8cm]{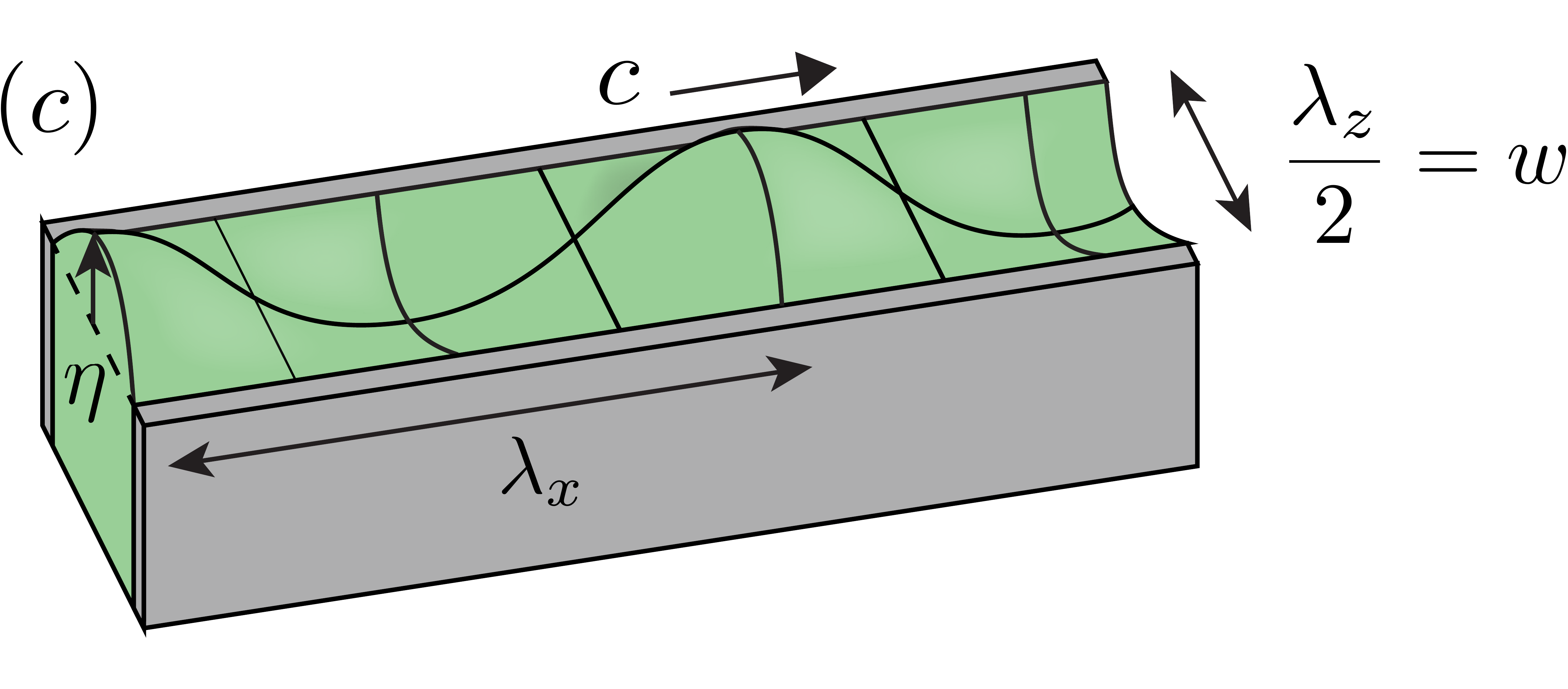}
      \addtocounter{subfigure}{1}
      \captionlistentry{}
      \label{fig:sketchWave}
  \end{subfigure}
  \end{minipage}
  \begin{subfigure}{0.50\textwidth}
      \centering
      \includegraphics[width=6.5cm]{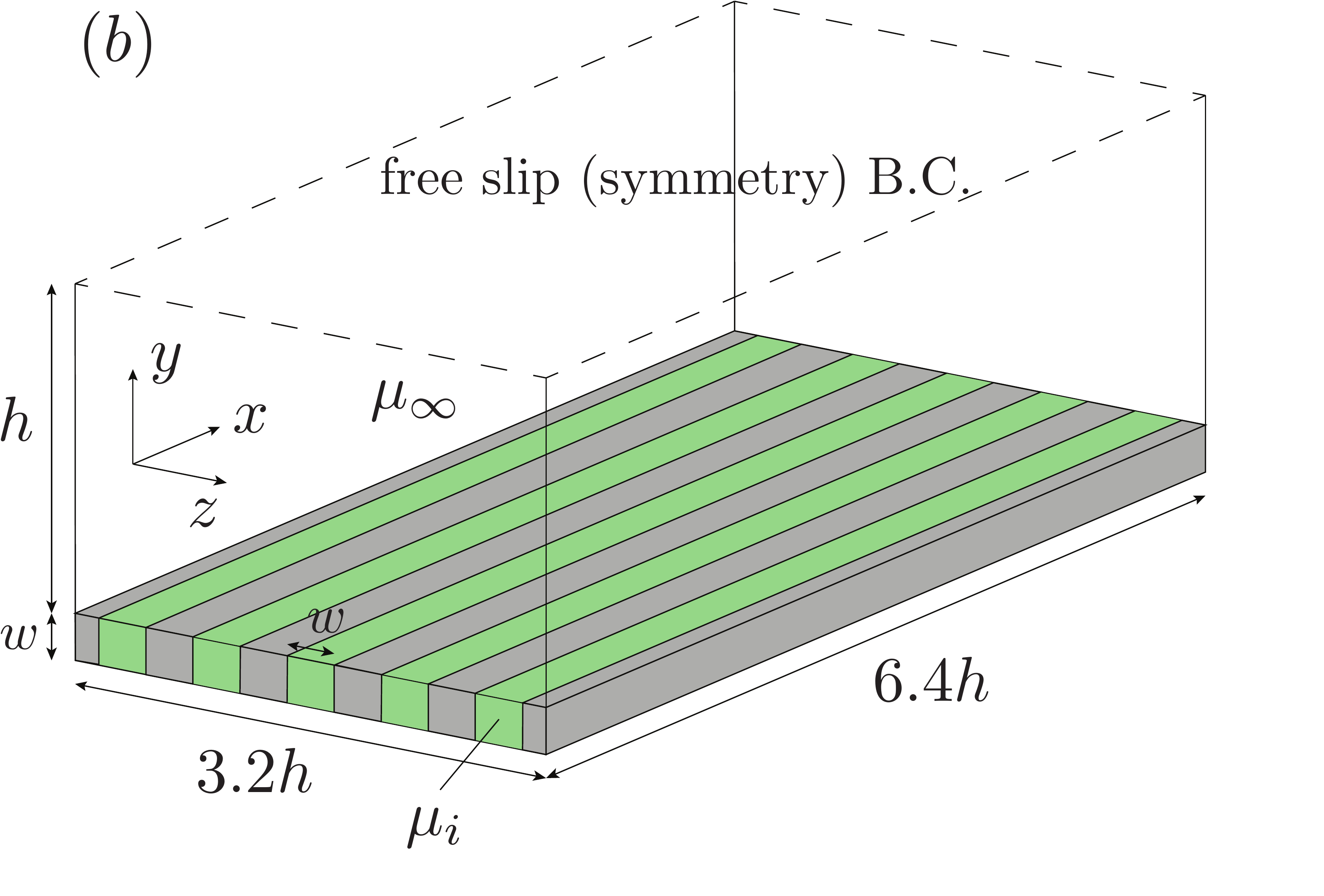}
      \addtocounter{subfigure}{-2}
      \captionlistentry{}
      \label{fig:sketch}
  \end{subfigure}
  \caption{($a$) Design map for LIS, spanned by $\WeNumber^+$ and $w^+$. Smooth and rough regions are separated using eqs.~\eqref{eq:min_phase_speed}, \eqref{eq:onset_condition_C} and \eqref{eq:onset_condition_C_slow}. Values from simulations with $w^{+0}=18$ 
  are included, with symbols referring to $\mu_i/\mu_\infty = 1$ (\opencirc), $\mu_i/\mu_\infty = 0.5$ (\opensquare) and $\mu_i/\mu_\infty = 2$ (\opendiamond) and colours to $\WeNumber = 50$ (green), $\WeNumber = 100$ (blue), $\WeNumber = 150$ (red), $\WeNumber = 200$ (yellow) and $\WeNumber = 400$ (turquoise). 
  Also shown are values with wider grooves, $w^{+0} = 36$, with $\WeNumber = 25$ (purple circle), $\WeNumber = 50$ (green circle) and $\WeNumber = 100$ (blue circle). 
  The asymptotic relations \eqref{eq:asymptotic_relation_small_w} and \eqref{eq:asymptotic_relation_large_w} are shown with (\longbroken) and (\chain), respectively.
  Sketch of ($b$) the channel configuration and ($c$) of a wave on a groove. The infused liquid is shown in green and the solids in grey.
  } 
  \label{fig:simulation_setup}
\end{figure}

The existence of energy transfer is not sufficient for failure of LIS, however.  The interface fluctuations also need to grow sufficiently fast to reach large amplitudes that induce roughness effects. Figure \ref{fig:stabilityMap}, which summarises our main contribution, shows three domains, namely rough, smooth and transitional (grey) in a plane spanned by a groove width ($w^+$) and a Weber number ($\WeNumber^+$), both normalised with the viscous length scale. This design map is obtained from the critical-layer theory and provides a means to design LIS that can be predicted to achieve a balance between performance (large $w^+$) and stability (smooth domain). 

\section{Numerical methods and configuration}
\label{sec:numerical_method}
We consider a fully developed turbulent open channel flow. The flow domain, shown schematically in  fig.~\ref{fig:sketch}, has the size $(L_x, L_y, L_z) = (6.4h, h + k, 3.2h)$, where $x,y$ and $z$ correspond to the streamwise, wall-normal and spanwise directions, respectively, and $h$ is the half-channel height. At the top boundary we impose a free-slip (symmetry) boundary condition (BC). Periodic boundary conditions are imposed in the streamwise and spanwise directions. The streamwise-aligned grooves at the bottom wall have a height $k$, a width $w$ and square cross-section, $k = w$. The fluid-solid ratio is set to $0.5$. The infused and external fluids have the same density $\rho_i=\rho_\infty=\rho$, but different viscosities ($\mu_i$ and $\mu_\infty$). We have used grooves of width $w^{+0} = 18$. Throughout this paper, $+0$ refers to normalisation using the friction velocity of a regular smooth wall (nominal wall units). A single superscript $+$ refers to normalisation using the friction velocity of each individual case. The corresponding viscous length scale is $\delta_\nu = \mu_\infty/(\rho u_\tau)$, where $u_\tau$ is the friction velocity.

We impose a constant mass flow rate through a uniform pressure gradient over $0 < y < h$, where $y = 0$ corresponds to the crest of the texture so that $\ReNumber_b = \rho h U_b/\mu_\infty=2820$, giving $\ReNumber_\tau  =\rho h u_\tau/\mu_\infty\approx 180$ (with the pressure gradient implemented as a volume force). Here, $\ReNumber_b$ and $\ReNumber_\tau$ are Reynolds numbers based on bulk velocity $U_b$ and friction velocity $u_\tau$, respectively. 

Our simulations allow for a moving  liquid-liquid-solid contact line with a dynamic contact angle different from the static value, which is $\theta = 45\degree$ with respect to the infused liquid. Our method also allows for interface deformation, which is typically quantified by the Weber number, defined as $\WeNumber = \rho U_\mathrm{b}^2 h/\gamma$, where $\gamma$ is the surface tension. We have simulated LIS for $\WeNumber = 100$, $150$ and $200$ and viscosity ratios $\mu_i/\mu_\infty = 0.5$, $1$ and $2$. The Weber number in wall units is $\WeNumber^+ = \rho u_\tau^2\delta_\nu/\gamma = \mu_\infty u_\tau/\gamma$. It can be noted that the Weber number based on the friction velocity and the width of the grooves is $\WeNumber^+ w^+$.

The numerical configuration described above corresponds to an infused liquid consisting of some alkane (with dynamic viscosities similar to that of water \citep{buren17}), a water channel with $h = 0.5$ cm and $U_\mathrm{b} = 1$ m/s. This results in $\ReNumber_b= 5000$, which is close to the value in our simulations. A typical surface tension $\gamma = 50$ mN/m then results in $\WeNumber = 100$. 

The code used for the simulations is based on the PArallel, Robust, Interface Simulator (PARIS), which employs a VOF method for the multiphase description \citep{aniszewski19, fuster18}. The cited papers also include additional test cases and validations. Height functions are used for curvature calculation for the surface tension. The interface is advected in the manner suggested by \citet{weymouth10} at each substep. This advection scheme conserves the volume of both liquids to a high accuracy. At solid surfaces, a contact angle is imposed by using the height functions and a dynamic contact angle model for VOF based on hydrodynamic theory \citep{legendre15}. Details are given in sec.~S1 of the supplementary material (SM). Finally, the grid size is $(N_x, N_y, N_z) = (256, 640, 1024)$, with constant grid spacing in each direction. The flow was converged after $600 h/U_b$, and statistics were collected over a time of $500 h/U_b$.

\section{Results}
\subsection{Dependence of drag on Weber number}
\label{sec:surface_tension}

\begin{figure}
    \centering
    \includegraphics[width=0.9\textwidth]{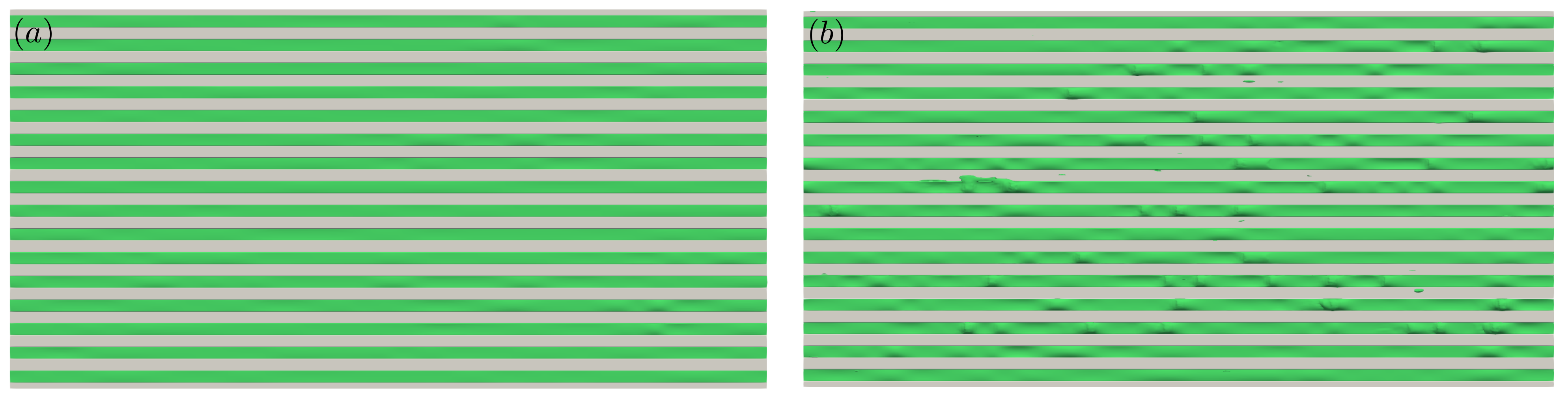}
    \caption{Top view of interfaces (green) and surface (grey) at one instant for $\mu_i/\mu_\infty = 1$ and ($a$) $\WeNumber = 100$ and ($b$) $\WeNumber = 200$. The flow is from left to right. The complete domain is shown.}
  \label{fig:interfaceTop}
\end{figure}

Figure \ref{fig:interfaceTop}(a) shows an instantaneous snapshot of the liquid-liquid interface, viewed from the top, for $\WeNumber = 100$ and $\mu_i = \mu_\infty$. There are oscillations on the interface, due to the finite surface tension, but they remain small. The deformation of the interface increases with $\WeNumber$, however. For $\WeNumber = 200$, significantly larger waves develop on the interface, as shown in fig.~\ref{fig:interfaceTop}(b).

The consequences of the waves on the overlying flow can be quantified by the drag reduction
\begin{equation}
    \DR = \frac{c_f^0 - c_f}{c_f^0},
    \label{eq:DR}
\end{equation} 
where $c_f = 2\tau_w/(\rho U_b^2)$ is the friction coefficient and $\tau_w$ is the total stress at the crest plane of the surface (computed from the pressure gradient). Here, $c_f^0$ is the  coefficient of a regular smooth wall at $y=0$. For $\WeNumber=100$ and $\WeNumber=200$, we obtained $\DR=0.09$ and $\DR=-0.04$, respectively. In other words, the capillary waves observed at $\WeNumber = 200$ increase frictional drag compared to a smooth and homogeneous surface and have therefore induced failure of the LIS.

\begin{figure}
  \centering
  \begin{subfigure}{0.45\textwidth}
      \centering 
      \includegraphics[width=5.5cm]{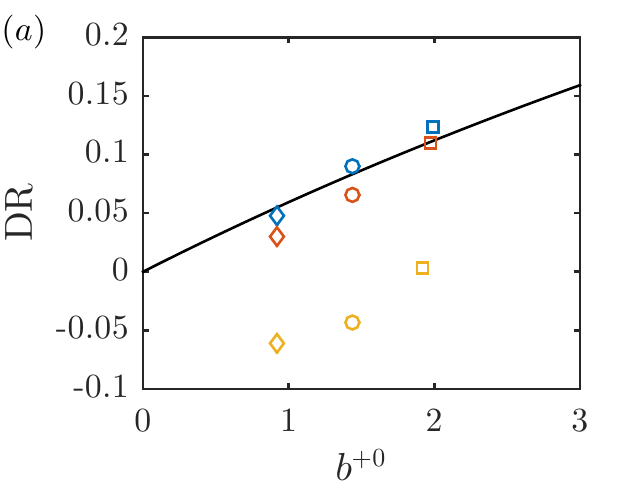}
      \captionlistentry{}
      \label{fig:slipLengthDR}
  \end{subfigure}
  \begin{subfigure}{0.45\textwidth}
      \centering
      \includegraphics[width=5.5cm]{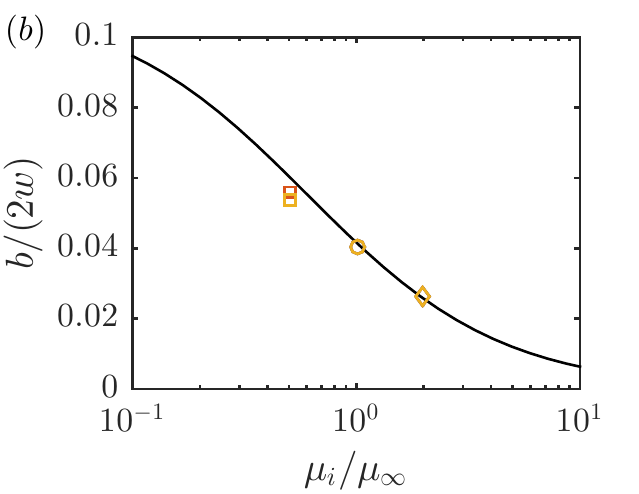}
      \captionlistentry{}
      \label{fig:ViscRslipLength}
  \end{subfigure}
  \caption{Drag reduction as a function of ($a$) the slip length and ($b$) slip length normalised by the pitch as a function of viscosity ratio. Symbols and colours are the same as in fig.~\ref{fig:stabilityMap}. In panel ($b$), the points for different $\WeNumber$ are almost on top of each other. Analytical relations are shown by black lines: ($a$) \full~\citet{rastegari15} and ($b$) \full~\citet{schonecker14}.}
  \label{fig:LISSlip}
\end{figure}

Figure \ref{fig:slipLengthDR} shows $\DR$ for  $\WeNumber=\{100,150,200\}$ and viscosity ratios $\mu_i/\mu_\infty = \{0.5,1, 2\}$ as a function of the apparent slip length, $b^{+0}$. The slip length $b$ is the distance at which the mean velocity would be zero if linearly extrapolated at the crests of the surface. 
It is largely unaffected by changes in $\WeNumber$, but it increases with decreasing viscosity ratio. In fact, the slip lengths extracted from our numerical simulations with $w^+ \approx 18$ are well approximated by the slip lengths obtained by solving the Stokes equations for a periodic array of grooves exposed to unit shear \citep{schonecker14} (black line in fig.~\ref{fig:ViscRslipLength}). A similar agreement was observed for smaller grooves ($w^+ \approx 9$) in turbulent flows by \citet{fu17, arenas19}. 

When the interface is perfectly flat ($\WeNumber=0$), the drag reduction can be related to slip length as
\begin{equation}
    \DR \approx \frac{b^{+0}}{b^{+0} + U_b^{+0}},
\label{eq:rasteg}
\end{equation}
where $U_b^{+0}$ is the bulk velocity in nominal wall units. This relation -- shown in \ref{fig:slipLengthDR} (black line) -- can be obtained 
by neglecting changes in the Reynolds shear stress above a smooth wall \citep{rastegari15}.
We observe from fig.~\ref{fig:slipLengthDR} that, for $\WeNumber = 100$ (blue) and $\WeNumber = 150$ (red), there is a drag reduction ($\DR>0$) for all three viscosity ratios. Moreover, the deviations from \eqref{eq:rasteg} are small, confirming that the drag reduction mechanism is indeed slippage. These small deviations are due to change of Reynolds shear stress. In contrast, the deviations from \eqref{eq:rasteg} are significant for $\WeNumber = 200$ (yellow), where we observe a drag increase ($\DR < 0$) for $\mu_i/\mu_\infty = 1$ and $2$ and a $\DR$ close to zero for $\mu_i/\mu_\infty = 0.5$. The corresponding mean velocity and velocity fluctuations reflect the increase in drag, and these are described in the SM (sec.~S2).

The waves formed on the interface at $\WeNumber = 200$ are sufficiently large to cause roughness effects. Interface height profiles at different times are shown in fig.~\ref{fig:waveWe200}, together with amplitudes of a wave in its initial stage (fig.~\ref{fig:waveAmplitude}, yellow). The wave amplitude is defined as the height of the local maximum of the wave. Wave amplitudes of $a^+ > 5$ are observed and these extend outside the viscous sublayer, indicating that the surface is transitionally rough. The amplitude grows initially at an exponential rate, before it levels off. In contrast, interface fluctuations for $\WeNumber=100$ have small amplitudes ($a^+<1$) (fig.~\ref{fig:waveWe100}) and show a significantly smaller growth rate (fig.~\ref{fig:waveAmplitude}, blue). The exponential growth rate (fig.~\ref{fig:waveAmplitude}, dashed) is an indication of a linear instability. In the next section, we provide evidence of a critical-layer instability \citep{miles57}, where energy is transferred to the wave perturbation from the turbulent flow.

\subsection{Conditions for phase speed and growth rate of capillary waves}
\label{sec:miles_instability}
We assume a small perturbation on the liquid-liquid interface of the form
\begin{equation}
    \eta = A e^{ik_x (x - ct)}\cos(k_z z),
    \label{eq:eta}
\end{equation}
where $z = 0$ is located in the centre of the groove. As illustrated in fig.~\ref{fig:sketchWave}, $\eta$ is the height of the interface, $A$ is the initial wave amplitude, $c$ is a complex wave speed and $t$ is time. Moreover, $k_x = 2\pi/\lambda_x$ and $k_z = 2\pi/\lambda_z$ are  streamwise and spanwise wavenumbers, respectively. The spanwise wavelength can have a maximum value of $\lambda_z = 2w$, due to the finite width of the grooves. This value can be seen to dominate in the snapshots of fig.~\ref{fig:interfaceTop}, as most waves only have one crest or one trough in the spanwise direction. We also observe from fig.~\ref{fig:interfaceTop} that streamwise wavelengths are generally similar to, or larger than, $\lambda_z$. This three-dimensionality implies that both spanwise and streamwise curvatures contribute to the capillary pressure of a wave, 
\begin{equation}
    \Delta p_\mathrm{cap} = p_0^+-p_0^-=\gamma\left(\frac{\partial^2}{\partial x^2} + \frac{\partial^2}{\partial z^2}\right)\eta = -\gamma k^2 \eta.
    \label{eq:capillaryPressure}
\end{equation}
Here, $k = \sqrt{k_x^2 + k_z^2}$ 
and $p_0^+$ ($p_0^-$) is the pressure above (below) the interface.

\begin{figure}
  \centering
  \begin{minipage}{0.59\textwidth}
  \begin{subfigure}{1.0\textwidth}
    \includegraphics[width=7.7cm]{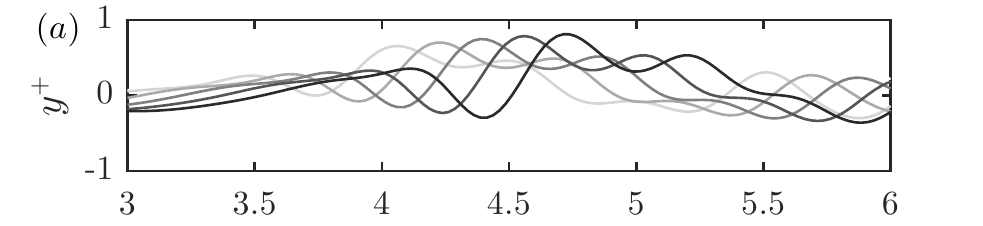}
    \captionlistentry{}
    \label{fig:waveWe100}
  \end{subfigure}
    \begin{subfigure}{1.0\textwidth}
    \includegraphics[width=7.7cm]{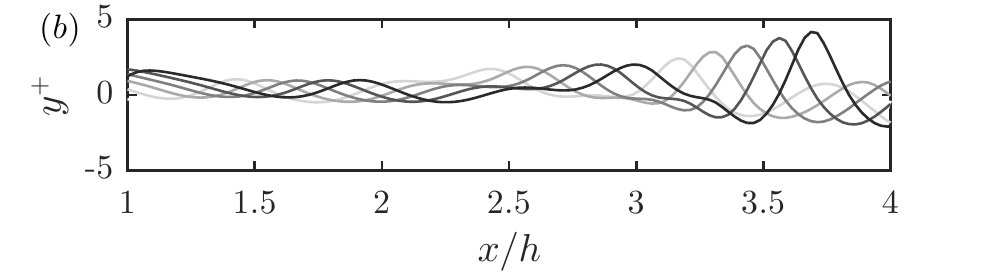}
    \captionlistentry{}
    \label{fig:waveWe200}
  \end{subfigure}
  \end{minipage}
  \begin{subfigure}{0.39\textwidth}
    \includegraphics[width=1.0\textwidth]{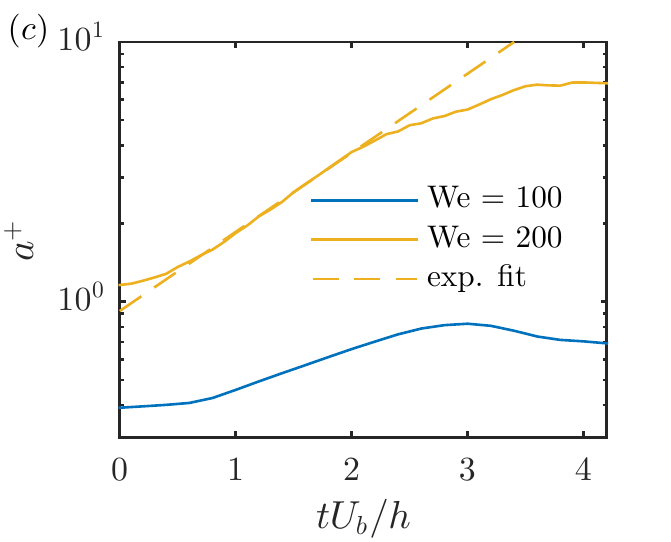}
    \captionlistentry{}
    \label{fig:waveAmplitude}
  \end{subfigure}
\vspace{-0.5cm}
  \caption{Instantaneous interface heights in the centerline of a groove for $\mu_i/\mu_\infty = 1$:  ($a$) $\WeNumber = 100$ for $3 \le x \le 6$ and ($b$) $\WeNumber = 200$ for $1 \le x \le 4$. The five profiles are separated by $\Delta t = 0.2 h/U_b$. Note the difference in  vertical scale. In ($c$), the wave amplitude developing at $x/h = 4.5$ in ($a$) and at $x/h = 3.5$ in ($b$) are shown. The phase speeds in ($a$) and ($b$) can be estimated as $c^+ \approx 14$ and $c^+ \approx 10$, respectively. }
  \label{fig:interfaceHeights}
\end{figure}

Next, we consider a wall-normal velocity disturbance $v(x,y,z,t)$ on the turbulent mean flow $U(y)$ with the same waveform as $\eta$.
If we neglect viscous and nonlinear effects, the amplitude $\hat v(y)$ is governed by the Rayleigh equation (SM, sec.~S3 A),
\begin{equation}
    \frac{1}{k^2}\hat v^{\prime\prime}-\left[1 + \frac{1}{(U-c)}\frac{1}{k^2} U^{\prime\prime}\right] \hat v = 0,
    \label{eq:rayleigh}
\end{equation}
where $^\prime$ denotes a derivative with respect to $y$. The velocity perturbation must vanish at infinity and satisfy the kinematic condition at the interface, $v/(U-c) = ik_x\eta$. The equation for the pressure amplitude, $\hat p(y)$,  corresponding to eq.~\eqref{eq:rayleigh} is
\begin{equation}
    \frac{\hat p}{\rho} = -i\frac{k_x}{k^2}\left [ (U-c)\hat v^{\prime} - U^{\prime}\hat v\right ].
    \label{eq:pressure}
\end{equation}
Our aim is to find an approximate solution to the equations (\ref{eq:capillaryPressure}-\ref{eq:pressure}) in order to determine the phase speed $\Re(c)$ (real part) and growth rate $\Im(k_xc)$ (imaginary part) of the interface perturbation \eqref{eq:eta}.

\citet{miles57} formulated a similar set of equations for describing wind-induced water waves, where gravity -- instead of capillarity -- balances fluid pressure. He suggested the following approximate solution for $v$:
\begin{equation}
    v = ik_x\eta(U-c)e^{-ky}, \qquad y\geq 0.
    \label{eq:vMiles}
\end{equation}
This expression, which satisfies the boundary conditions at $y\rightarrow \infty$, implies that $1/k$ is the relevant length scale over which $v$ decreases. The assumption of exponential decay can also be used in the grooves:
  \begin{equation}
    v = - ik_x c\eta e^{ky}, \qquad y<0.
    \label{eq:vGroove}
\end{equation}
Here, we have assumed that the grooves are sufficiently deep such that the velocity perturbation is nearly zero at the bottom of the groove. With a depth $w$, $kw > k_zw \ge (2\pi/(2w)) w = \pi$, and, since $e^{-\pi} \ll 1$, the assumption is valid for our configuration.  We have also neglected $U$ and its derivative inside the groove. Inserting \eqref{eq:vGroove} into \eqref{eq:pressure}
results in the following expression for the  pressure immediately below the interface ($y\rightarrow 0^-$): %
\begin{equation}
    p_0^{-} = \rho\frac{k_x^2}{k}c^2\eta.
    \label{eq:groovePressure}
\end{equation}

Similarly, by inserting \eqref{eq:vMiles} into \eqref{eq:pressure} the pressure just above  the interface is 
\begin{equation}
    p_0^+ = (\alpha + i\beta)\rho U_1^2 \frac{k_x^2}{k} \eta, 
    \label{eq:milesPressure}
\end{equation}
where $\alpha$ and $\beta$ are real constants and $U_1$ is an arbitrary reference velocity. It is shown in SM (sec.~S3 C) that the parameter $\alpha$ can be decomposed into two parts, $\alpha = \alpha_1 + \alpha_2$, where $\alpha_1$ corresponds to eq.~\eqref{eq:groovePressure} and $\alpha_2$ incorporates the remaining contributions from the slip velocity and the shear. Using this decomposition and inserting \eqref{eq:groovePressure} and \eqref{eq:milesPressure} into \eqref{eq:capillaryPressure}, we obtain (SM, sec.~S3 D),
\begin{equation}
    c = c_w\left(1 + \frac{1}{4}(\alpha_2 + i\beta)\frac{U_1^2}{c_w^2} + \dots\right).
    \label{eq:eigenvalue}
\end{equation}
Here $c_w = \sqrt{\gamma k^3/(2\rho k_x^2)}$ is the free phase speed, i.e. the speed of a capillary wave without forcing from the overlying flow.

\begin{figure}
    \centering
    \begin{subfigure}{0.45\textwidth}
      \centering
      \includegraphics[width=5.5cm]{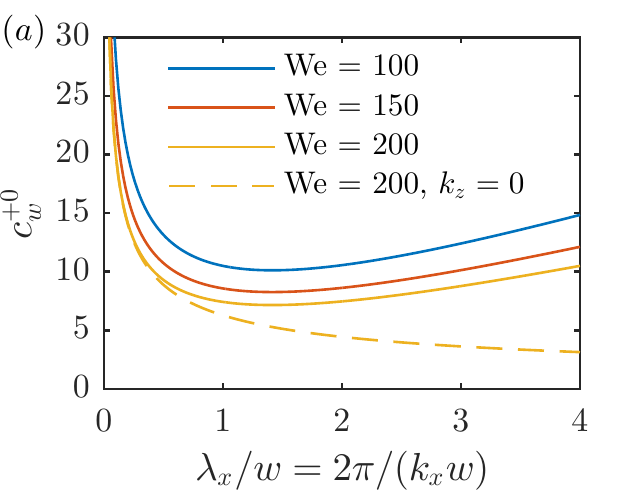}
       \captionlistentry{}
      \label{fig:phaseSpeeds}
    \end{subfigure}
    \begin{subfigure}{0.45\textwidth}
      \includegraphics[width=5.5cm]{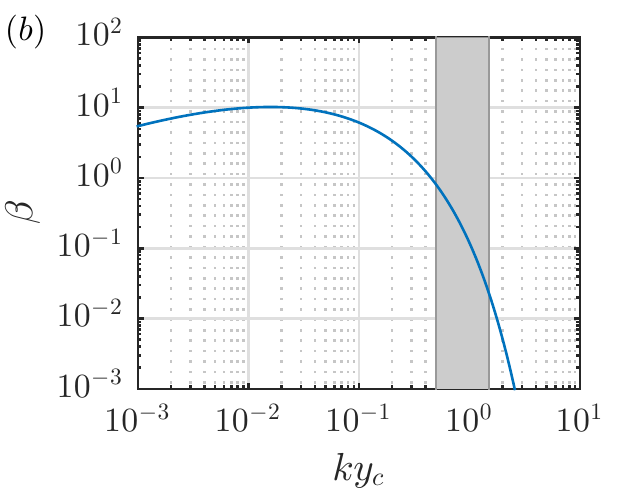}
      \captionlistentry{}
      \label{fig:beta}
    \end{subfigure}
    \caption{($a$) The free phase speed $c^{+0}_w$ for $\WeNumber = \{100, 150, 200\}$ when $k_z = \pi/w$ (i.e. $\lambda_z/w = 2$). For $\WeNumber = 200$, the phase speed of a two-dimensional wave ($k_z = 0$) is also shown. ($b$) The growth rate coefficient $\beta$ versus $ky_c$, showing a fast decrease in $\beta$ for $ky_c \gtrsim 1$.
    }
    \label{fig:stability}
\end{figure}

The free phase speed is shown in fig.~\ref{fig:phaseSpeeds} (in nominal wall units) as a function of $\lambda_x/w$ for $\WeNumber=\{100,150,200\}$. We note that two-dimensional capillary waves ($k_z = 0$)  have a phase speed that monotonically decreases with $\lambda_x$.  However, for LIS, there is a minimum  phase speed due to the finite spanwise wavelength. This minimum is approximately (for the analytical expression see SM, sec.~S3 D)
\begin{equation}
    c_{w,\mathrm{min}}^+ \approx \sqrt{\frac{\pi}{\WeNumber^+ w^+}}.
    \label{eq:min_phase_speed}
\end{equation}
For $\WeNumber = 200$ (and $\mu_i/\mu_\infty = 1$)¸ $c_{w,\mathrm{min}}^+ = 7.0$, which is slightly lower than the phase speed of the wave shown in fig.~\ref{fig:waveWe200}. One may use $c^+_{w,\min}$ as a lower bound of the actual phase speed of capillary waves on LIS. 

We now turn our attention to the imaginary part of \eqref{eq:eigenvalue} to approximate the growth rate of the instability. As shown by \cite{miles57} -- and repeated in the SM (sec S3 E) -- one may integrate the Rayleigh equation and evaluate the pressure equation \eqref{eq:pressure} at the interface to find 
\begin{equation}
    \beta = -\pi\left|\frac{v_c}{k_x\eta U_1}\right|^2\frac{1}{k}\frac{U_c''}{U_c'},
     \label{eq:beta}
\end{equation}
where the subscript $c$ denotes values at the position of the critical layer $y_c$. In order for an infinitesimal wave to have a positive growth rate, i.e.~$\beta > 0$, a first requirement is that $U''_c<0$, i.e. negative curvature at the critical layer. This is satisfied if the critical layer is outside of the viscous sublayer.

A second requirement for $\beta>0$  is that $v_c$ in eq.~\eqref{eq:beta} is non-zero at the critical layer. The approximate solution of $v$ in eq.~\eqref{eq:vMiles}
implies, however, that $v$ is zero at the critical layer. As shown in SM (sec.~3 E), one may transform the condition for positive growth rate to an integral form to estimate $v$ in the vicinity of the critical layer. By further assuming a logarithmic mean velocity profile and setting the reference velocity to $U_1=u_\tau/\kappa$ (where $\kappa$ is the von K\'arm\'an constant), one may evaluate the expression for $\beta$ (as a function of $ky_c$), and obtain what is shown in fig.~\ref{fig:beta}. 

We observe from fig.~\ref{fig:beta} that, when $ky_c > 3/2$, then $\beta < 0.02$, which results in very slow-growing waves, whereas when $ky_c < 1/2$, we have $\beta>0.8$, resulting in a factor 40 or more faster growth. The grey region in fig.~\ref{fig:beta} marks the range $1/2 < ky_c < 3/2$ where there is a transition from low to high growth rates. Now, since $k>k_z$ and the upper limit of $\lambda_z$ is  $2w$, we may formulate bounds for the position of the critical layer. When
\begin{equation}
 y_{c}^+ \lesssim  \frac{1}{2}\frac{w^+}{\pi},
    \label{eq:onset_condition_C}
\end{equation}
the growth rate can be expected to be significant, in contrast to when
\begin{equation}
    y_{c}^+ \gtrsim  \frac{3}{2}\frac{w^+}{\pi},
    \label{eq:onset_condition_C_slow}
\end{equation}
for which the growth is negligible.

Equations \eqref{eq:min_phase_speed}, \eqref{eq:onset_condition_C} and \eqref{eq:onset_condition_C_slow} provide relationships between $\WeNumber^+, w^+$  and $c_{w,\min}^+,y_{c,\max}^+$ that can be confirmed by our simulations. Equation \eqref{eq:min_phase_speed} states that a large $\WeNumber^+$ and/or $w^+$ give a small phase speed. This is observed qualitatively by following the travelling waves on the interface in fig.~\ref{fig:interfaceHeights}(a,b). More quantitatively, the space-time correlations of the interface height for $\WeNumber = 100$ and $\WeNumber = 200$ give $c^+ = 15.1$ and $c^+ = 10.5$, respectively (SM fig.~S8).

Compared to $\WeNumber=100$, the lower phase speed for $\WeNumber = 200$ results in a lower position of the critical layer. When the height of the critical layer approaches the interface and satisfies eq.~\eqref{eq:onset_condition_C}, the growth rate coefficient $\beta$ of the waves (eq.~\ref{eq:beta}) is large. This is confirmed by our simulations, where we observe in fig.~\ref{fig:waveAmplitude} that both the growth rate and interface amplitudes are larger for $\WeNumber=200$ compared to $\WeNumber=100$.

We use an inviscid model here to get a tractable analytical solution, and to illustrate the important physics involved. It has been shown that the effect of introducing viscosity on capillary waves with relevant wavenumbers would be a slight damping \citep{jeng98}. However, it is possible that viscosity influences the velocity induced by the waves deep inside the grooves to a higher extent. 

\subsection{Implications for the design of LIS}
\label{sec:implications}
The conditions \eqref{eq:min_phase_speed}, \eqref{eq:onset_condition_C}  and \eqref{eq:onset_condition_C_slow} can be used as design criteria for LIS. 
One may expect a high-performing LIS by choosing a groove width and a surface tension of the infused liquid such that -- for relevant friction Reynolds numbers -- the design falls within the smooth region of fig.~\ref{fig:stabilityMap}. This region is defined by $(\WeNumber^+,w^+)$, where $y^+_c\geq 1.5 w^+/\pi$, and thus from eq.~\eqref{eq:onset_condition_C_slow} very 
small growth rates of capillary waves are predicted.
Conversely, the rough region in fig.~\ref{fig:stabilityMap} shows $(\WeNumber^+,w^+)$, where $y^+_c\leq 0.5 w^+/\pi$, and therefore waves will amplify rapidly. Here, we may expect either a very low-performing LIS or even a drag-increasing LIS due to roughness effects. In between the smooth and rough domains, we show in fig.~\ref{fig:stabilityMap} a transitional region (grey), which corresponds to $0.5w^+/\pi\leq y^+_c\leq 1.5w^+/\pi$. Here, we cannot predict if the resulting waves induce roughness effects using our analytical approach. It should be mentioned that the boundaries of the transitional region in fig.~\ref{fig:stabilityMap} are determined in three steps: (i) given $w^+$, determine $y^+_c$ from \eqref{eq:onset_condition_C} (lower boundary) or \eqref{eq:onset_condition_C_slow} (upper boundary); (ii) given $y^+_c$, determine $c^+$ (phase speed) from $U^+(y^+_c)=c^+$, where $U^+(y)$ is a turbulent mean profile of a smooth wall; and finally (iii) given $c^+$, assume $c^+\approx c^{+0}_{w,\min}$ and determine $\WeNumber^+$ from \eqref{eq:min_phase_speed} (or the exact coefficient of \eqref{eq:min_phase_speed} given in the SM). 

Figure \ref{fig:stabilityMap} also shows scaling laws between smooth and rough regions. For small $w^+$ (and thus $y_c^+$), we may assume that the critical-layer velocity is $U_c^+ \approx y_c^+$. This is acceptable right above the viscous sublayer where the mean flow has some curvature. Then eq.~\eqref{eq:min_phase_speed} gives that the height of the lowest critical layer is $y_c^+ \approx \sqrt{\pi/(\WeNumber^+w^+)}$. By assuming that $y_c^+\sim w^+/\pi$, we obtain 
\begin{equation}
    w^+\sim (\WeNumber^+)^{-1/3},
    \label{eq:asymptotic_relation_small_w}
\end{equation} 
which is shown with dashed line in fig.~\ref{fig:stabilityMap}. It is observed that this asymptotic relation represent a reasonable scaling law for $w^+\lesssim20$.

For larger $w^+$, away from the viscous sublayer, we assume $U^+ = (1/\kappa) \log(y^+) + B$, where $B$ is a constant. This gives a nonlinear relation
\begin{equation}
     1/(\sqrt{\WeNumber^+w^+}) \sim (1/\kappa) \log(w^+/\pi).
         \label{eq:asymptotic_relation_large_w}
\end{equation}
This curve is shown in fig.~ \ref{fig:stabilityMap} with a dashed-dotted line, and provides a scaling of the neutral curve for $w^+\gtrsim 30$.

The scaling laws illustrate that, when $w^+$ increases,  there needs to be rapid decrease of $\WeNumber^+$ to remain in the smooth region. For example, for $w^+\approx 70$, we need $\WeNumber^+ \approx 3\cdot 10^{-4}$, which corresponds to $\WeNumber \approx 10$. This is relevant for drag reduction, since the width of the grooves should be maximised for a given surface tension to optimise $\DR$ (see fig.~\ref{fig:LISSlip}), but without entering the rough zone in fig.~\ref{fig:stabilityMap}. Note that for a fixed geometry, increasing the flow speed, and thereby $u_\tau$, increases both $\WeNumber^+$ and $w^+$, so that the design needs to be made for the largest flow speed to which the surface is exposed.

Finally, in fig.~\ref{fig:stabilityMap}, the values of our numerical simulations are shown with symbols. These also include more extreme Weber numbers, $\WeNumber = 50$ and $\WeNumber = 400$ (using $\mu_i/\mu_\infty = 1$), which resulted in a drag reduction of $9.3\%$ and $-15\%$, respectively, confirming the trend of the other simulations. In addition to the simulations at $w^{+0}=18$, we also show points (circles) for larger grooves of width $w^{+0} = 36$ (also using $\mu_i/\mu_\infty = 1$). For these grooves, there was a drag reduction by $18\%$ for $\WeNumber = 25$ (purple circle) and $17\%$ for $\WeNumber = 50$ (green circle), whereas for $\WeNumber = 100$ (blue circle), the drag reduction was lowered to $2\%$ and we observed large waves. This implies that the growth rate rapidly increases between the last two cases as they fall in the transitional zone in fig.~\ref{fig:stabilityMap}.

\begin{figure}
  \centering
    \begin{subfigure}{0.45\textwidth}
      \centering
      \includegraphics[width=4.3cm]{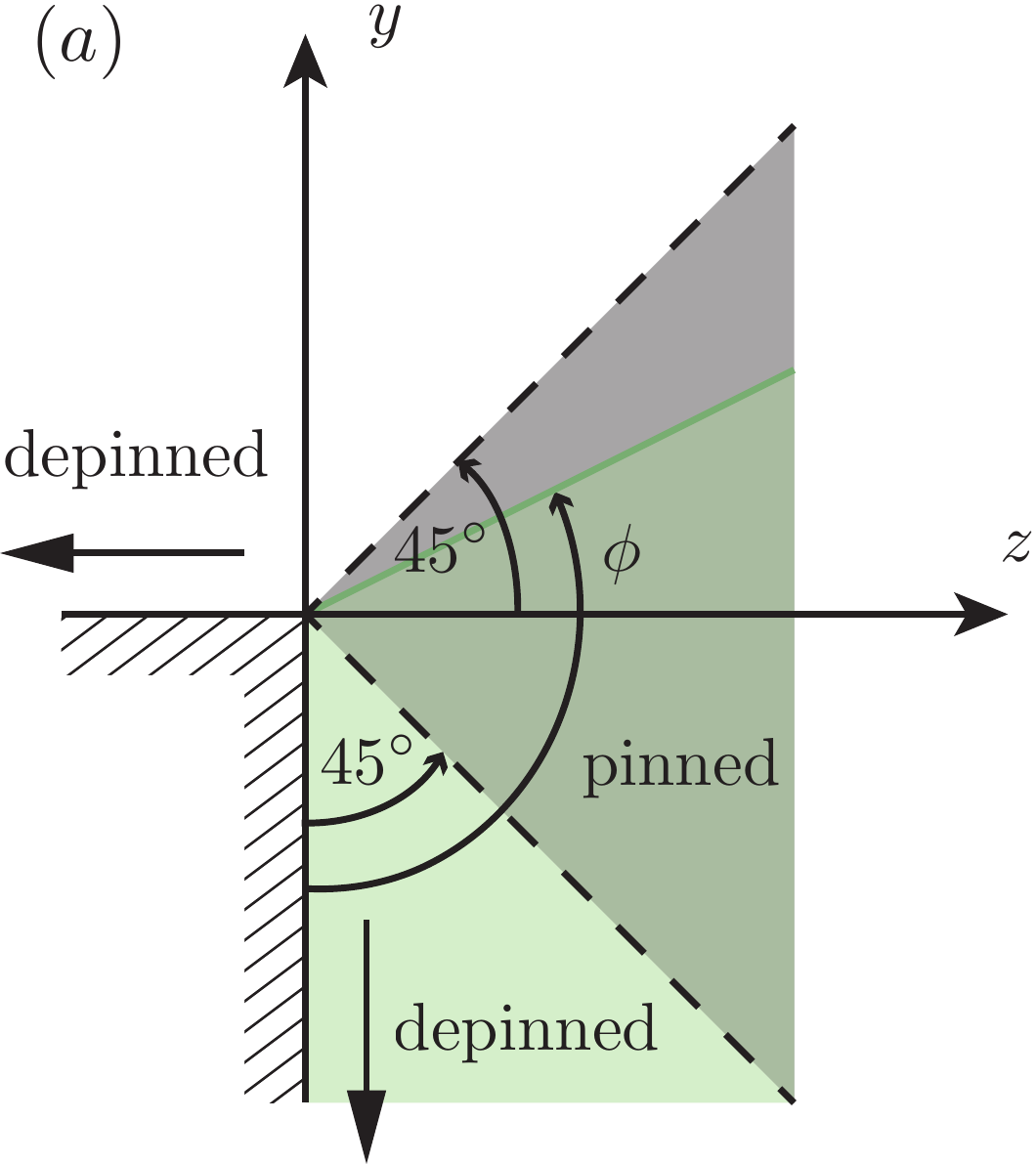}
      \captionlistentry{}
      \label{fig:phiIllustration}
  \end{subfigure}
  \begin{subfigure}{0.45\textwidth}
      \centering
      \includegraphics[height=4.2cm]{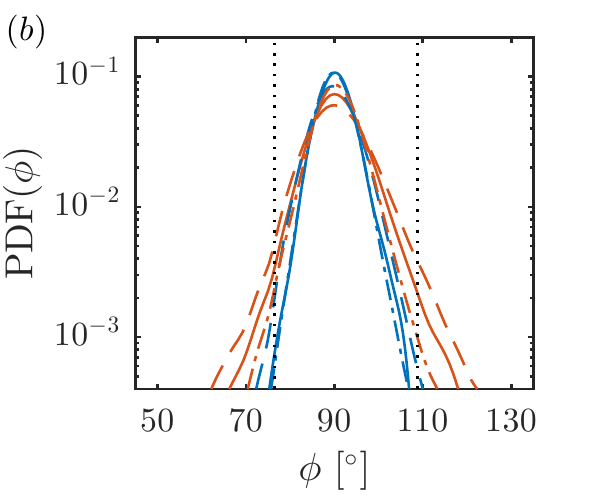}
      \captionlistentry{}
      \label{fig:phiDistLog}
  \end{subfigure}
  \caption{($a$) If $\theta < \phi < \theta + 90\degree$, the contact line remains pinned according to Gibbs' criterion (grey area). This is illustrated for $\theta = 45\degree$. If $\phi$ is outside this range, the contact line depins, and moves in the direction indicated by the arrows. ($b$) The PDF of $\phi$ from simulations for the pinned cases with $w^{+0} = 18$, $\theta = 45\degree$, $\mu_i/\mu_\infty = 0.5$ (\longbroken), $\mu_i/\mu_\infty = 1$ (\full) and $\mu_i/\mu_\infty = 2$ (\chain) and the Weber numbers $\WeNumber = 100$ (blue) and $\WeNumber = 150$ (red). The boundaries of the interval corresponding to a probability of 95\% for the widest PDF are also shown (\dotted).
  }
  \label{fig:gibbs_criterion}
\end{figure}

\subsection{Contact line depinning}
The capillary waves modify the contact angle between the interface and the wall, and may potentially result in a depinning of the interface from the corners of the ridges. 
According to Gibbs' criterion, which is a purely geometrical criterion, the interface remains pinned if $\theta < \phi < 90\degree + \theta$, where $\phi$ is the angle the interface makes to the inner wall of the groove. The lower limit is the limit for when the contact line moves into the groove, while the upper is the limit for when it moves on top of the ridge \citep{gibbs06}. This is illustrated in fig.~\ref{fig:phiIllustration}.  In contrast to the cases $\WeNumber = 100$ or $150$ (fig.~\ref{fig:interfaceTop}), we observed that for $\WeNumber = 200$ the interface depinned occasionally due to  the waves on the interface.

The measured probability density functions (PDF) of $\phi$ for $\WeNumber = 100$ and $150$ with $\theta = 45\degree$ are plotted in fig.~\ref{fig:phiDistLog} for all three viscosity ratios. Since the contact line was observed to remain pinned for these parameters, the PDF are independent of $\theta$ and can be used to predict limits for the contact angle. It can be noted that the PDF for all parameters shown are centred around $\phi = 90\degree$ and that $\phi$ is unlikely to reach below $70\degree$ or above $110\degree$. The interval between these values corresponds to a probability of more than 95\% for the widest PDF. The standard deviation of $\phi$ decreases with $\mu_i/\mu_\infty$ and increases with $\WeNumber$, as is indicated by the width of the PDF. This is to be expected, since the dissipation rate increases with the viscosity and the restoring force of surface tension becomes weaker with increasing $\WeNumber$.

A restoring force for the contact line also comes from mass conservation. If the contact line occasionally does depin into the groove on one position, it will be raised elsewhere. Based on this observation and the statistics in fig.~\ref{fig:phiDistLog}, depinning is not the main failure mode of LIS for the geometry chosen in this study. Depinning is, however, expected to be important for a LIS with grooves of finite length.

\section{Conclusions}
\label{sec:conclusions} 
We have explored the behaviour of LIS in a turbulent channel flow with square longitudinal grooves for $\ReNumber_\tau\approx 180$. By allowing the interface and the contact line to move, we could investigate the unconstrained motion of the interface. 
For a fixed groove width, we found a rapid increase in drag of LIS above a certain Weber number due to the appearance of large capillary waves. The generation mechanism of these waves was elucidated using the theory developed by \citet{miles57}. The limit for when these waves act as roughness is set by the width of the grooves $w^+$ and the Weber number $\WeNumber^+$, as illustrated in fig.~\ref{fig:stabilityMap}. It should also be noted that these non-dimensional parameters depend on the flow speed. Using an analytical analysis, we have provided scaling laws and design criteria for robust drag-reducing LIS. Specifically, the relations show how to achieve a balance between large groove widths (enhancing drag reduction) and high surface tension of the infused liquid (enhancing stability) for different flow speeds.

\backsection[Acknowledgements]{This work was supported by SSF, the Swedish Foundation for Strategic Research (Future Leaders grant FFL15:0001). Simulations were performed on resources provided by the Swedish National Infrastructure of Computing (SNIC).}

\backsection[Declaration of interests]{The authors report no conflict of interest.}

\backsection[Supplementary data]{Supplementary material are available at \\https://doi.org/10.1017/jfm.2021.241. \\}


\bibliographystyle{jfm} \bibliography{references}
\end{document}


\title{Supplementary Material: Roughness on liquid-infused surfaces induced by capillary waves}

\author{Johan Sundin\aff{1}, Stéphane Zaleski\aff{2,3} and Shervin Bagheri\aff{1} }
\affiliation{\aff{1}Linné FLOW Centre, Dept.~Engineering Mechanics, KTH, Stockholm, Sweden \\\aff{2}Sorbonne Universit\'e \& CNRS, Institut Jean Le Rond d'Alembert, UMR 7190, Paris, France \\\aff{3}Institut Universitaire de France, Institut Jean Le Rond d’Alembert, UMR 7190, Paris, France}

\maketitle

\section{Details of numerical methods}
\label{sec:app_numerical methods}
In this section, we describe the schemes of the numerical code that was used to solve the two-component flow over and inside the textured surfaces. For momentum convection, a second-order central scheme was used, whereas a third-order Runge-Kutta scheme was used for time integration \citep{costa18}. The equation for the pressure was solved with a fast Fourier transform (FFT) solver (FFTW) and handling domain decomposition with the 2DECOMP\&FFT library. To describe solids, we used a grid-aligned immersed-boundary method (IBM) \citep{breugem05}. With this method, the edge of the solid is located at the edge of the cells. For a staggered grid, this means that the top of the solid ridges are at the location of the wall-normal velocity nodes. 

To impose a contact angle at the liquid-liquid-solid contact line, we specified the height function of the first ghost layer \citep{afkhami08}. Since the IBM is grid-aligned, the implementation of the contact angle is the same on an the immersed boundaries and the domain boundaries. We used a dynamic model of the contact angle based on hydrodynamic theory \citep{blake06} -- adapted to VOF \citep{legendre15} -- , together with a no-slip velocity condition. As shown in \citep{legendre15}, the dynamic contact angle also improves grid independence. The method is explained further in sec.~\ref{sec:dynamic_contact_angle}. It should be noted that the ridge corners' ghost cells were set to always be interface cells, and that the contact angle was imposed when the interface moved to adjacent cells. Depinning occured for the higher Weber numbers.

\begin{figure}[H]
    \begin{subfigure}{0.32\textwidth}
        \includegraphics[width=5.5cm]{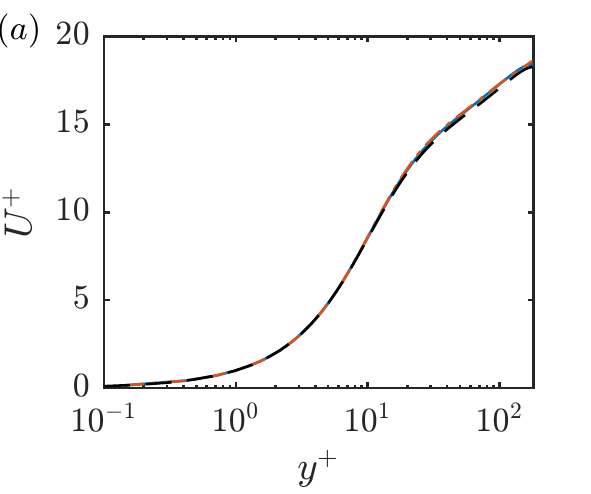}
        \captionlistentry{}
        \label{fig:meanVelocity}
    \end{subfigure}
    \begin{subfigure}{0.32\textwidth}
        \includegraphics[width=5.5cm]{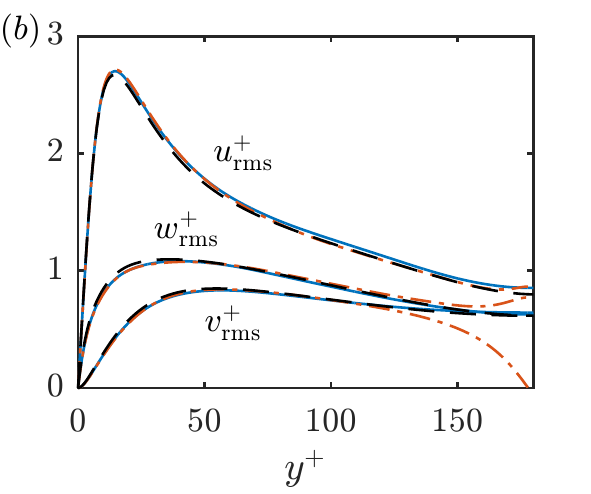}
        \captionlistentry{}
        \label{fig:rmsVelocities}
    \end{subfigure}
    \begin{subfigure}{0.32\textwidth}
        \includegraphics[width=5.5cm]{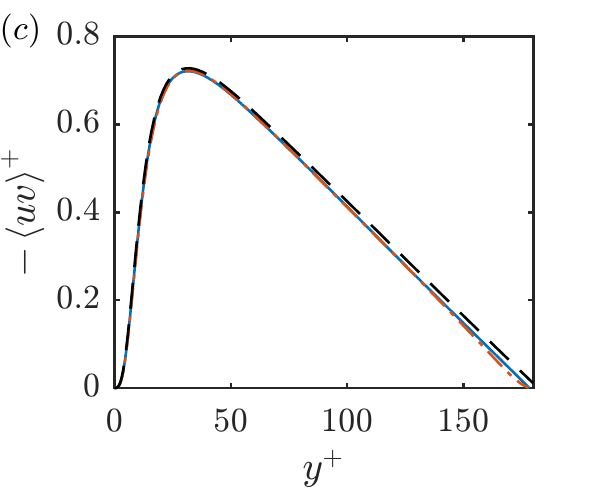}
        \captionlistentry{}
        \label{fig:reynoldsShearStress}
    \end{subfigure}
    \caption{Simulation of a turbulent channel flow with smooth walls at $\ReNumber_\tau \approx 180$, with mean velocity ($a$), r.m.s.~velocities ($b$), and Reynolds shear stress ($c$). The spanwise velocity r.m.s.~value is represented by $w_\mathrm{rms}^+$ for convenience, but elsewhere $w$ refers to the groove width. \full~PARIS; \chain~PARIS open channel; \longbroken~Lee \& Moser (2015) \cite{lee15}}
    \label{fig:smoothWallValidation}
\end{figure}

The turbulent flow was validated by comparing the mean flow and the r.m.s.~velocities for a full channel with smooth walls to data from ref.~\cite{lee15}. The results are shown in fig.~\ref{fig:smoothWallValidation}, having for the mean velocity a deviation of $0.8$\% at the channel center. Shown are also statistics of a smooth open channel, with only small deviations for $y^+ < 100$.

\begin{figure}[H]
  \centering
  \includegraphics[width=7cm]{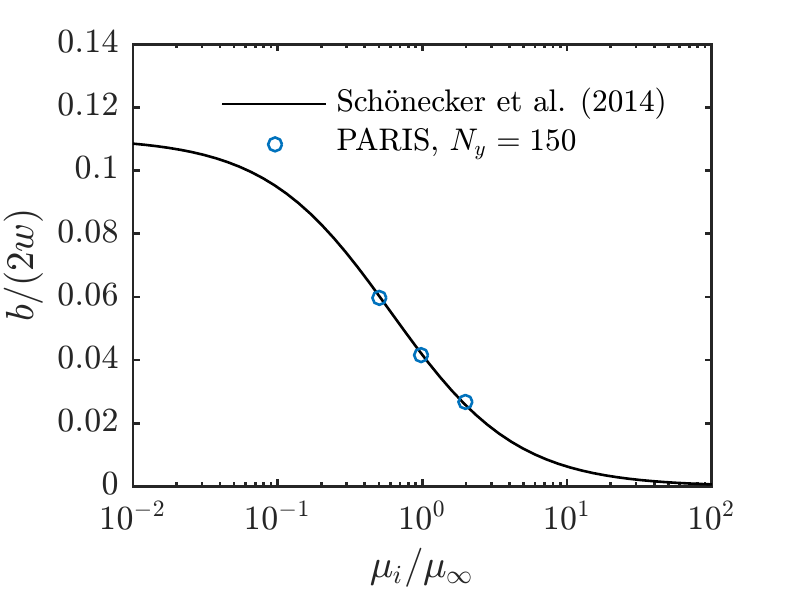}
  \caption{Comparison of slip length with the expressions of Schönecker \textit{et~al.} (2014). $N_y$ refers to the total number of grid cells in the wall-normal direction, with $N_y/3 = 50$ in the groove.}
  \label{fig:schonecker}
\end{figure}

The description of the streamwise liquid filled grooves was validated against the analytical expressions of Sch\"onecker et al.~\cite{schonecker14} for the slip length in a laminar flow. These expressions have recently also been used as a reference for turbulent data of LIS simulations \cite{fu17}. For these tests, we use a computational box corresponding to a unit cell of the surface, with height three times the height of the groove. On the top boundary, a constant shear was applied in the streamwise direction. A similar setup was recently used to investigate the robustness of LIS with spanwise grooves \cite{ge18}. Having 50 cells in the grooves, the errors were less than $4\%$ for $\mu_i/\mu_\infty = 0.5, 1$ and $2$, see fig.~\ref{fig:schonecker}. 

\begin{figure}[H]
    \begin{subfigure}{0.32\textwidth}
        \includegraphics[width=5.5cm]{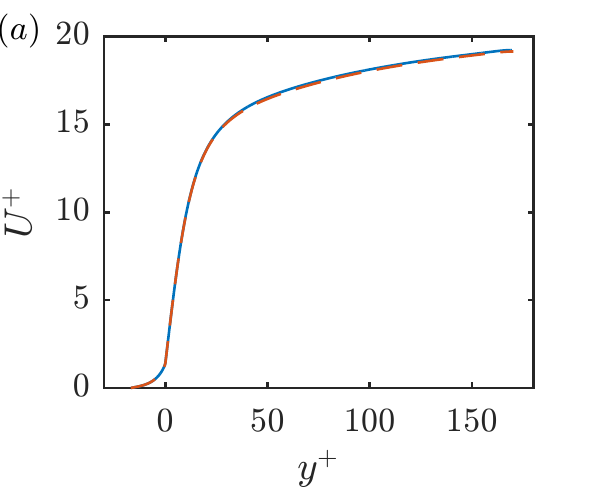}
        \captionlistentry{}
    \end{subfigure}
    \begin{subfigure}{0.32\textwidth}
        \includegraphics[width=5.5cm]{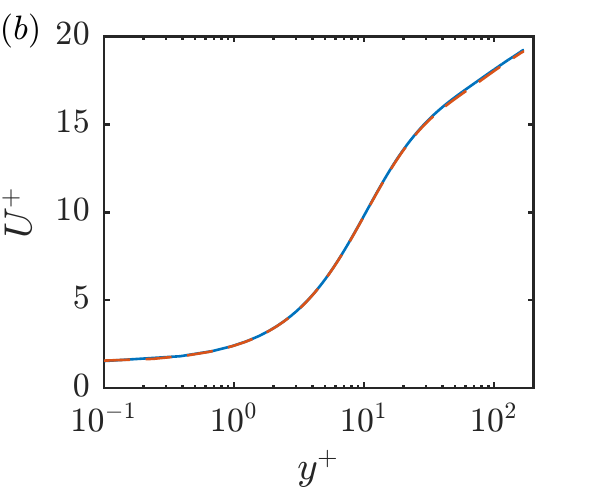}
        \captionlistentry{}
    \end{subfigure}
    \begin{subfigure}{0.32\textwidth}
        \includegraphics[width=5.5cm]{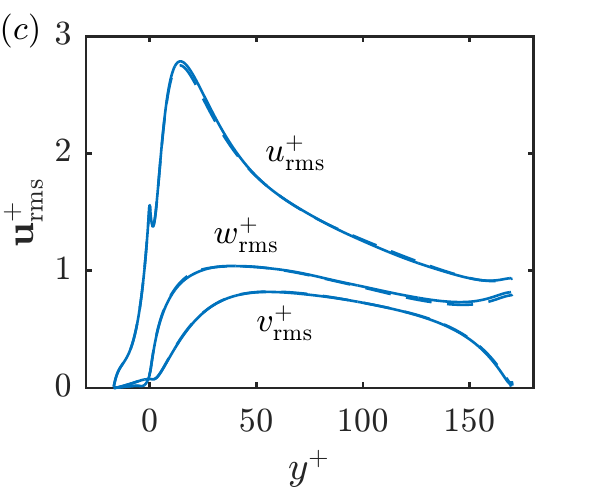}
        \captionlistentry{}
    \end{subfigure}
    \caption{Grid refinement study: \full~regular grid; \longbroken~refined grid. The spanwise velocity r.m.s.~value is represented by $w_\mathrm{rms}^+$ for convenience.}
    \label{fig:LISValidation}
\end{figure}

A grid refinement study was also performed for $\mu_i/\mu_\infty = 1$, $\WeNumber = 100$ and $\theta = 45 \degree$, where another grid with $(N_x, N_y, N_z) = (384, 960, 1536)$ was used, increasing the total number of grid cells by a factor of 3.4. This gave a change of the friction coefficient by 0.76\%. Plots are shown in fig.~\ref{fig:LISValidation}.

\subsection{Dynamic contact angle}
\label{sec:dynamic_contact_angle}

Following ref.~\cite{legendre15}, the contact angle can be found from the equation
\begin{equation}
    G^*(\theta_\mathrm{num}) = G^*(\theta_\mathrm{stat}) + \CaNumber\ln\left(\frac{\Delta/2}{\lambda} \right),
    \label{eq:coxAngle}
\end{equation}
where $\theta_\mathrm{num}$ is the numerically imposed contact angle, $\theta_\mathrm{stat}$ is the static angle, $\CaNumber = U_\mathrm{cl}\sqrt{\mu_i\mu_\infty}/\gamma$, with contact line velocity $U_\mathrm{cl}$, $\lambda$ is a microscopic length scale corresponding to an effective slip length of the contact line, here taken to be constant, and $\Delta$ is the wall-normal cell height. The angle $\theta_\mathrm{stat}$ is in this manuscript denoted only $\theta$ for simplicity. The contact line velocity is measured half a cell above the wall. The function $G^*(\theta)$ is a monotonically increasing function. It is defined as $G^*(\theta) = \sqrt{q}G(\theta)$, where $q = \mu_\infty/\mu_i$ and $G(\theta)$ is the original function derived by ref.~\cite{cox86}, with the extra factor for increased symmetry,
\begin{equation}
    G^*(\theta) = \int^\theta_0 f^{*-1}(\theta', q)\dif \theta',
\end{equation}
with
\begin{equation}
    f^{*-1}(\theta, q) = \frac{q^{0.5}(\theta^2 - \sin^2\theta)[(\pi - \theta) + \sin\theta\cos\theta] + q^{-0.5}((\pi - \theta)^2 - \sin^2\theta)[\theta - \sin\theta\cos\theta]}{2\sin\theta[(2 - q - q^{-1})\sin^2\theta + q\theta^2 + q^{-1}(\pi - \theta)^2 + 2\theta(\pi - \theta)]}.
\end{equation}
When written in this form, it is apparent that $f^*(\theta, q) = f^*(\pi - \theta, q^{-1})$, which is a necessary requirement, since eq.~\eqref{eq:coxAngle} should be independent of which of the two components we consider. This is identical to the hydrodynamic theory of an apparent angle model suggested in ref.~\cite{cox86}, where $\theta_\mathrm{num}$ is the apparent angle. Legendre and Maglio~\cite{legendre15} has suggested that $\lambda$ should be on the order of the the physical slip length (typically on the order of $1$ nm), and may or may not be used in combination with a numerically applied slip length. We use a no-slip condition, with a numerical value of $\lambda = 2\cdot10^{-7}h$, that with $h \approx 0.5$ cm attains a realistic value. The full equation \eqref{eq:coxAngle} was solved by the Newton's method, and $f^{*-1}(\theta, N)$ was integrated with the trapezoidal rule. For numerical reasons, the value has been limited to $30\degree \le \theta_\mathrm{num} \le 150\degree$.

\begin{figure}
  \centering
  \begin{subfigure}{0.45\textwidth}
      \centering
      \includegraphics[width=6cm]{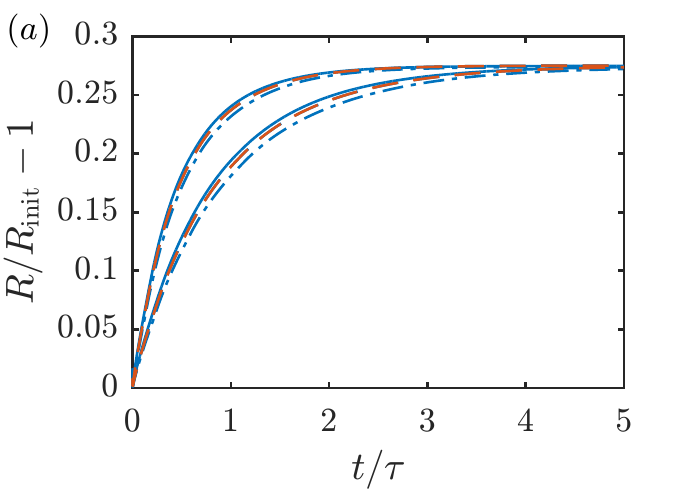}
      \captionlistentry{}
      \label{fig:dropletSpreading}
  \end{subfigure}
  \begin{subfigure}{0.45\textwidth}
      \centering
      \includegraphics[width=6cm]{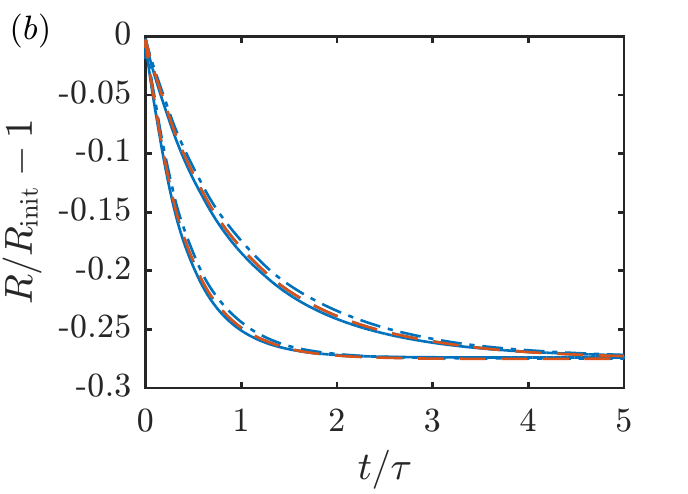}
      \captionlistentry{}
      \label{fig:dropletContraction}
  \end{subfigure}
  \caption{Radius at wall ($y = 0$) of droplet spreading from $90\degree$ to $60\degree$ ($a$), and contracting from $90\degree$ to $120\degree$ ($b$), for three different wall-normal cell sizes: \full~$N_y = 32$; \longbroken~$N_y = 64$; \chain~$N_y = 128$, in blue. The viscosity ratios are $0.5$ and $2$, with the faster spread/contraction for the case with surrounding fluid of lower viscosity. Data from a droplet on an immersed boundary are also shown, using the intermediate resolution (red, \longbroken).}
  \label{fig:dropletTest}
\end{figure}

We evaluated the grid-independence of the contact line model by looking at droplet spreading for three different cell sizes. A three dimensional half-sphere droplet of radius $R_\mathrm{init} = 0.5$ was placed in a box of size $(L_x, L_y, L_z) = (2,1,2)$, with periodic boundary conditions in the $x$- and $z$-directions, and no-slip and shear-free boundary conditions in the negative and positive $y$-direction, respectively. The initial radius sets the length scale of the problem. The contact angle was initially $90\degree$, when the static contact was set to $\theta_\mathrm{stat} = 60\degree$, so that the droplet started to spread. The droplet had a density $\rho=1$, viscosity $\mu = 0.25$ and surface tension $\gamma = 7.5$. Using the density and the viscosity we can define a time scale $\tau = \rho R_\mathrm{init}^2/\mu$. The surrounding fluid had the same density, and the viscosity ratio was changed by varying the viscosity of the surrounding fluid. Viscosity ratios of both $0.5$ and $2$ were tested, to span the range of this study. The number of cells of the wall parallel directions were $N_x = 64$ and $N_z = 64$, and the number of cells of the wall-normal direction was varied between $N_y = 32$, $N_y = 64$ and $N_y = 128$. The microscopic length scale was set to $\lambda = 2\cdot10^{-7}$. Resulting spreading radii are shown in fig.~\ref{fig:dropletSpreading}. The differences between the curves are a few percentage, considered small enough for this study. Corresponding droplet contraction tests, with contraction from $90\degree$ to $120\degree$, are shown in fig.~\ref{fig:dropletContraction}. The tests using the intermediate resolution were also repeated on an immersed boundary by extending the domain in the $y$-direction below the droplet by $\Delta L_y = 0.5$ and adding a solid slab there. The curves are hardly distinguishable between the two setups, as shown in fig.~\ref{fig:dropletTest}. The fluid parameters and the dynamic contact angle model are the same as those used by Legendre \& Maglio \cite{legendre15} (Dyn2) for the spreading case, except for the viscosity ratio (there equal to 1) and a slightly different value of $\lambda$. Their results agree well with those shown here.

\section{Turbulence statistics}
\label{sec:turbulence_statistics}
\begin{figure}[H]
  \centering
  \begin{subfigure}{0.45\textwidth}
      \centering
      \includegraphics[width=5.5cm]{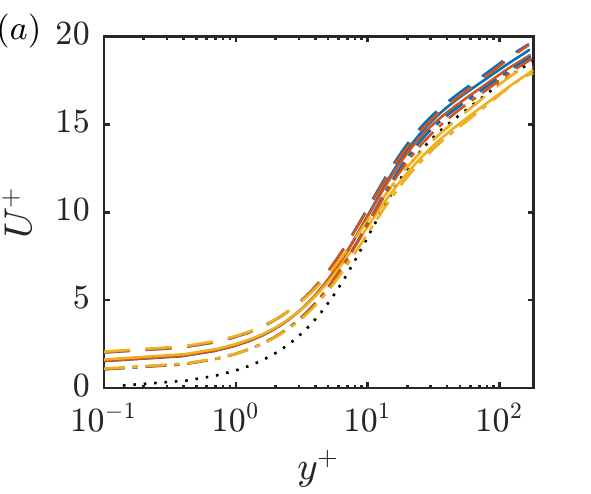}
       \captionlistentry{}
      \label{fig:meanVelocityLIS}
  \end{subfigure}
  \begin{subfigure}{0.45\textwidth}
      \centering
      \includegraphics[width=5.5cm]{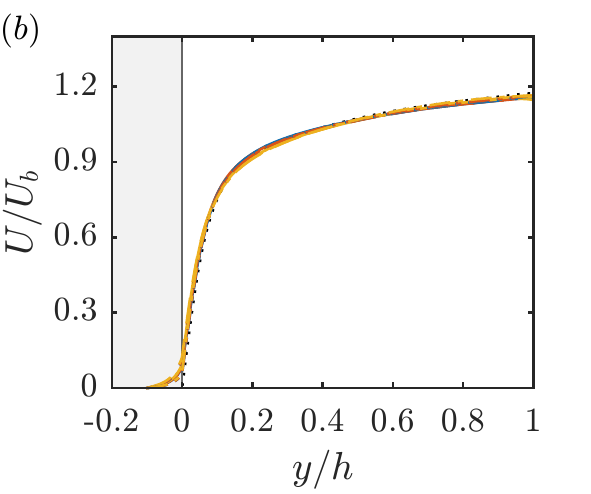}
       \captionlistentry{}
      \label{fig:meanVelocityOuterLIS}
  \end{subfigure}
    \begin{subfigure}{0.45\textwidth}
        \centering
        \includegraphics[width=5.5cm]{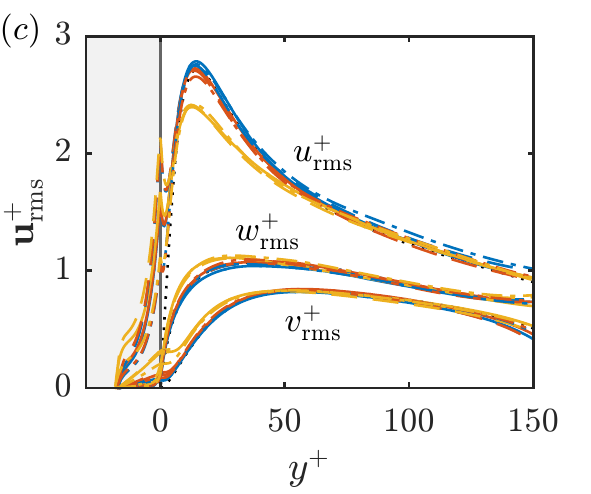}
         \captionlistentry{}
        \label{fig:rmsVelocitiesLIS}
    \end{subfigure}
    \begin{subfigure}{0.45\textwidth}
        \centering
        \includegraphics[width=5.5cm]{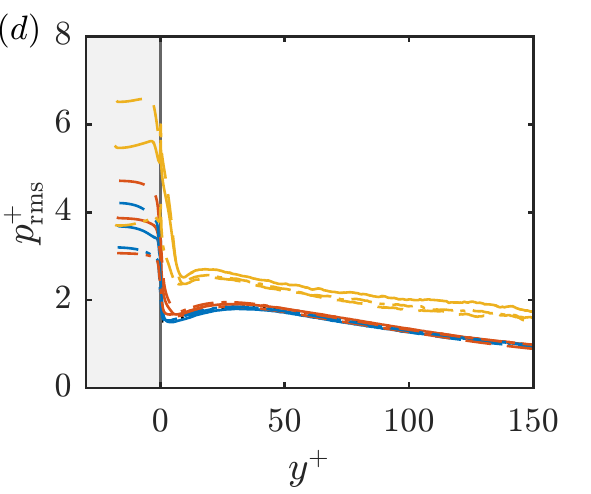}
        \captionlistentry{}
        \label{fig:rmsPressureLIS}
    \end{subfigure}
  \caption{Mean velocity profiles in wall units ($a$), and outer units ($b$) together with r.m.s.~velocities ($c$) and r.m.s.~pressure ($d$). The grey area represent the grooves. Colours represent different $\WeNumber$ as in fig.~3 ($\WeNumber = 100$-blue, $\WeNumber = 150$-red and $\WeNumber = 200$-yellow), and the different lines represent different viscosity ratios, $\mu_i/\mu_\infty = 0.5$ (\longbroken), $\mu_i/\mu_\infty = 1$ (\full) and $\mu_i/\mu_\infty = 2$ (\chain). Also shown are statistics for a smooth wall (\dotted). The spanwise velocity r.m.s.~value is represented by $w_\mathrm{rms}^+$ for convenience, but elsewhere $w$ refers to the groove width. }.
  \label{fig:LISStatistics}
\end{figure}

In this section, we show the turbulent statistics corresponding to the simulations $\WeNumber=\{100,150,200\}$ and viscosity ratios $\mu_i/\mu_\infty=\{0.5,1,2\}$.
Mean velocity profiles are plotted in fig.~\ref{fig:meanVelocityLIS} in wall units and fig.~\ref{fig:meanVelocityOuterLIS} in outer units. As can be seen from the plot in wall units, the centerline velocity is increased by the LIS for low $\WeNumber$. This is related to the drag reduction achieved at these $\WeNumber$, whereas the opposite occurs for the cases of drag increase \cite{jimenez04}. A reduction of the centerline velocity is followed by a decrease in the streamwise r.m.s.~component, and an increase of the wall-normal and the spanwise fluctuations, shown in fig.~\ref{fig:rmsVelocitiesLIS}. Over all, drag increase results in an increased isotropy of the velocity fluctuations. Increased wall-normal velocity fluctuations at the surface has a strong connection to the increase in drag \cite{orlandi03, orlandi06}. It implies stronger flow ejections, caused by the roughness the waves impose. 

The pressure r.m.s.~profiles are shown in fig.~\ref{fig:rmsPressureLIS}. Outside the grooves, the profiles for $\WeNumber = 100$ and $\WeNumber = 150$ are similar to the smooth wall profile, but are higher inside the grooves. For $\WeNumber = 200$, when there are large waves, the pressure fluctuations are increased also outside the grooves. This could be due to the increased roughness, but also due to formation of droplets.

\begin{figure}[H]
  \centering
  \includegraphics[width=6cm]{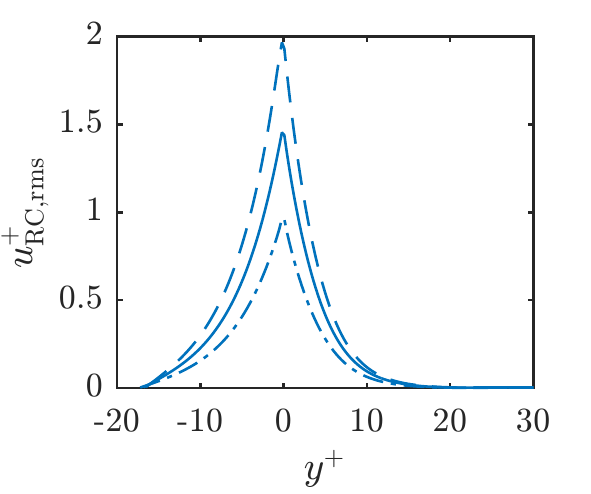}
  \caption{Streamwise dispersive stress for $\WeNumber = 100$, $\theta = 45\degree$ and $\mu_i/\mu_\infty = 0.5$ (\longbroken), $\mu_i/\mu_\infty = 1$ (\full) and $\mu_i/\mu_\infty = 2$ (\chain).}
  \label{fig:roughnessComp}
\end{figure}

In the streamwise r.m.s.~component, fig.~\ref{fig:rmsVelocitiesLIS}, there is a peak at $y = 0$, but not for the spanwise nor for the wall-normal. This can be related to the dispersive stresses from the solid structures, which are the r.m.s.~of the roughness-coherent velocity, $\mathbf{u}_\mathrm{RC}$. Due to the symmetry in the streamwise direction, the roughness-coherent velocity is the velocity field averaged in the streamwise direction and time. A peak is then created because the streamwise velocity component on average is larger over the interface at the grooves than over the ridges. The peak height differs between the viscosity ratios, but not so much between the different Weber numbers. This reflects the behaviour of the mean velocity at the interface, fig.~\ref{fig:meanVelocityLIS} for small $y^+$, as well as the slip length, fig.~3b. It was seen that both the spanwise and wall-normal roughness-coherent components were close to zero. The streamwise part is shown in fig.~\ref{fig:roughnessComp} for $\WeNumber = 100$.

\section{The Miles instability on grooves}
\subsection{Governing equations}
In this section, we look at the governing equations for the Miles instability. We start from the equations for a infinitesimal disturbance on a baseflow $U = U(y)$ 
\cite{schmid12}, 
\begin{alignat}{2}
    &\frac{\partial u}{\partial t} + U\frac{\partial u}{\partial x} + vU' &= -\frac{1}{\rho}\frac{\partial p}{\partial x}, \\
    &\frac{\partial v}{\partial t} + U\frac{\partial v}{\partial x} &= -\frac{1}{\rho}\frac{\partial p}{\partial y}, \label{eq:wall-normalMomentum}\\
    &\frac{\partial w}{\partial t} + U\frac{\partial w}{\partial x} &= -\frac{1}{\rho}\frac{\partial p}{\partial z},
\end{alignat}
and
\begin{equation}
        \frac{\partial u}{\partial x} + \frac{\partial v}{\partial y} + \frac{\partial w}{\partial z} = 0,
        \label{eq:continuity}
\end{equation}
where $'$ denotes a derivative in the $y$-direction. The spanwise velocity component is here represented by $w$ for convenience, which elsewhere refer to the groove width. Eliminating $p$ gives Rayleigh's equation for $v$ (inviscid Orr-Sommerfeld equation),
\begin{equation}
    \left[\left(\frac{\partial}{\partial t} + U\frac{\partial}{\partial x}\right)\nabla^2 - U''\frac{\partial}{\partial x}\right]v = 0,
    \label{eq:rayleigh}
\end{equation}
with $p$ given by 
\begin{equation}
    \frac{1}{\rho}\left[\frac{\partial^2}{\partial x^2} + \frac{\partial^2}{\partial z^2}\right]p = \frac{\partial^2 v}{\partial t\partial y} + U\frac{\partial^2 v}{\partial x\partial y} - U'\frac{\partial v}{\partial x}.
    \label{eq:pressure}
\end{equation}

We further assume an interface shape of a wave travelling in the streamwise direction,
\begin{equation}
    \eta = A e^{ik_x (x - ct)}\cos(k_z z) = A e^{k_xc_it}e^{ik_x (x - c_rt)}\cos(k_z z) = a(t)e^{ik_x (x - c_rt)}\cos(k_z z),
\end{equation}
where $\eta$ is the location of the interface, $A$ is the wave amplitude, $k_x$ is the streamwise wavenumber, $k_z$ is the spanwise wavenumber and $c = c_r + ic_i$. Henceforth, we'll use the notation $k = \sqrt{k_x^2 + k_z^2}$. The curvature of the interface gives rise to a pressure difference over the surface of (going from negative to positive $y$),
\begin{equation}
    \Delta p_\mathrm{cap} = \gamma\left(\frac{\partial^2}{\partial x^2} + \frac{\partial^2}{\partial z^2}\right)\eta = -\gamma k^2 \eta.
    \label{eq:capillaryPressure}
\end{equation}

The perturbation must die out at infinity,
\begin{equation}
    v \rightarrow 0 \quad \text{ as } \quad y \rightarrow \infty.
    \label{eq:infinityBC}
\end{equation}
In addition to this, fluid parcels on the interface must remain on the interface, (so that the interface remains a streamline) \cite{acheson91},
\begin{equation}
    \frac{\partial \eta}{\partial t} + U\frac{\partial \eta}{\partial x} = v \implies \frac{v}{U-c} = ik_x\eta \quad \text{ on $y = \eta \approx 0$}.
    \label{eq:interfaceKinematics}
\end{equation}

Returning now to Rayleigh's equation, eq.~\eqref{eq:rayleigh}, with $(x,z,t)$-dependence of $v$ as $\eta$,
\begin{equation}
    (U-c)\frac{\partial^2 v}{\partial y^2} - [k^2(U-c) + U'']v = 0.
    \label{eq:rayleighWave}
\end{equation}
From eq.~\eqref{eq:pressure}, with $(x,z,t)$-dependence of $p$ as $\eta$,
\begin{equation}
    \frac{p}{\rho} = -i\frac{k_x}{k^2}\left[(U-c)\frac{\partial v}{\partial y} - U'v\right].
    \label{eq:pressureWave}
\end{equation}

\subsection{Non-dimensionalisation}
In order to facilitate the derivations, we introduce the transform,
\begin{equation}
    \xi = ky, \quad U-c = U_1\tilde{u}(\xi) \quad \text{and} \quad v = ik_xU_1\tilde{v}(\xi)\eta,
\end{equation}
where $U_1$ is an arbitrary reference velocity. From the kinematics of the interface, eq.~\eqref{eq:interfaceKinematics}, it follows that 
\begin{equation}
    \tilde{v}_0 = \tilde{u}_0,
    \label{eq:interfaceKinematicsCondition}
\end{equation}
where a subscript $0$ denote the location of the interface, $y \approx 0$. The Rayleigh equation becomes
\begin{equation}
    \tilde{v}''\tilde{u} - [\tilde{u} + \tilde{u}'']\tilde{v} = 0.
    \label{eq:uvODE}
\end{equation}

The equation for the pressure, eq.~\eqref{eq:pressureWave}, results in
\begin{equation}
    p = \rho U_1^2 \frac{k_x^2}{k} \eta (\tilde{u}\tilde{v}' - \tilde{u}' \tilde{v}).
    \label{eq:pressureNonDim}
\end{equation}
The pressure just above the interface can be written as
\begin{equation}
    p_0^+ = (\alpha + i\beta)\rho U_1^2 \frac{k_x^2}{k} \eta.
    \label{eq:milesPressure}
\end{equation}
By comparing these two expressions,
\begin{equation}
    \alpha + i\beta = (\tilde{u}_0\tilde{v}_0' - \tilde{u}_0' \tilde{v}_0) = \tilde{u}_0\left(\tilde{v}_0' - \tilde{u}_0'\right),
    \label{eq:alphaBetaCondition}
\end{equation}
where the last equality comes from eq.~\eqref{eq:interfaceKinematicsCondition}. Following ref.~\cite{miles57}, $\tilde{u}$ is assumed to be approximately real and thus $\beta$ must come from the imaginary part of $\tilde{u}_0\tilde{v}_0'$, whereas $\alpha$ is the remaining part.

\subsection{Derivation of $\alpha$}
We here derive an expression for $\alpha$, which is the real part of eq.~\eqref{eq:alphaBetaCondition},
\begin{equation}
    \alpha = \Re(\tilde{u}_0\tilde{v}_0') - \tilde{u}_0\tilde{u}_0'.
\end{equation}
We use an approximate solution suggested by Miles \cite{miles57}, $\tilde{v} = \tilde{u}(\xi) e^{-\xi}$ (satisfying the boundary conditions), instead of solving the full equation. This approximate solution gives
\begin{equation}
    \alpha = -\tilde{u}_0^2 - \tilde{u}_0\tilde{u}_0' = \frac{1}{U_1^2}\left(-(U_0-c)^2 - \frac{U'_0}{k}(U_0-c)\right).
\end{equation}
Now, $\alpha$ can be decomposed into two parts, $\alpha = \alpha_1 + \alpha_2$, where $\alpha_1 = -c^2/U_1^2$ and $\alpha_2$ is the remaining part. We assume that $U_0$ can be neglected in comparison to $c$. The inverse of the wave number has an upper bound,
\begin{equation}
    \frac{1}{k} = \frac{1}{\sqrt{k_x^2 + k_z^2}} < \frac{1}{k_z} = \frac{\lambda_z}{2\pi} < \frac{2w}{2\pi},
    \label{eq:wavenumberLimit}
\end{equation}
since $k_x > 0$ and $\lambda_z$ must be smaller than two groove widths. Therefore,
\begin{equation}
    \frac{U'_0}{k} = \frac{1}{k h} \frac{\ReNumber_\tau^2}{\ReNumber_b}U_b = \frac{11}{k h} U_b < \frac{11}{\pi}\frac{w}{h}U_b,
\end{equation}
for $\ReNumber_\tau = 180$. For this Reynolds number and groove width $w^+ = 18$, the limiting value is $5.5$ in wall units. The shear could hence potentially influence the phase speed with up to this magnitude, however we consider it here to be of secondary importance.

We now look at the flow inside the groove. The wall-normal velocity must be zero at the bottom at the groove,
\begin{equation}
    \tilde{v} = 0 \quad \text{ at } \quad \xi = -kw.
\end{equation}
It can also be noticed that $kw > k_zw \le 2\pi/(2w) w = \pi$, and $e^{-\pi} = 0.043 \ll 1$. Hence, if the same assumption of exponential decrease of $\tilde{v}$ is made inside the groove, it is reasonable. We therefore assume $\tilde{v} = \tilde{u}e^{\xi}$ for $y < 0$. Neglecting also the mean velocity and its derivative inside the groove, the pressure is
\begin{equation}
    p^{-} = \rho\frac{k_x^2}{k}c^2\eta e^{ky}, \quad y < 0,
    \label{eq:groovePressure}
\end{equation}
as given by eq.~\eqref{eq:pressure} or eq.~\eqref{eq:pressureNonDim}. This is the pressure corresponding to $\alpha_1$.

\subsection{Phase speed and minimum phase speed}
The difference between the pressure above (eq.~\ref{eq:milesPressure}) and below (eq.~\ref{eq:groovePressure}) the interface must be balanced by the capillary pressure (eq.~\ref{eq:capillaryPressure}),
\begin{equation}
    \Delta p = p_0 - p^{-}(y=0) = (\alpha + i\beta)\rho U_1^2 \frac{k_x^2}{k} \eta - \rho\frac{k_x^2}{k}c^2 \eta = \Delta p_\mathrm{cap} = - \gamma k^2 \eta.
\end{equation}
Solving for $c$ and including the expression for $\alpha_1$,
\begin{equation}
    2c^2 = \frac{\gamma}{\rho}\frac{k^3}{k_x^2} + (\alpha_2 + i\beta)U_1^2 \implies c = c_w\left(1 + \frac{1}{4}(\alpha_2 + i\beta)\frac{U_1^2}{c_w^2} + \dots\right),
\end{equation}
where $c_w = \sqrt{\dfrac{\gamma}{2\rho}\dfrac{k^3}{k_x^2}}$ is the unforced phase speed neglecting any contributions from $\alpha_2$ and $\beta$ and the dots denote higher order terms. 

It is possible to find the $k_x$ for which $c_w$ assumes a minimum value by evaluating $\dif c_w/\dif k_x = 0$. The minimum value is, for $k_z = 2\pi/(2w)$,
\begin{equation}
    c_{w, \mathrm{min}} = \sqrt{\dfrac{\gamma}{\rho}\frac{3^{3/2}}{4}k_z} = \sqrt{\dfrac{\gamma}{\rho}\frac{3^{3/2}}{4}\frac{\pi}{w}} \implies c_{w, \mathrm{min}}^+ = \sqrt{\frac{1}{\WeNumber^+ w^+}\frac{3^{3/2}\pi}{4}} \approx \sqrt{\frac{\pi}{\WeNumber^+ w^+}}.
\end{equation}

\subsection{Derivation of $\beta$}
In this section, we find an expression for $\beta$. Returning to eq.~\eqref{eq:uvODE}, dividing by $\tilde{u}$, multiplying by the complex conjugate of $\tilde{v}$, namely $\tilde{v}^*$, integrating from $\xi = \xi_0$ to $\xi = \infty$, integrating $\tilde{v}''\tilde{v}^*$ by parts, and using the boundary conditions above (eqs.~\ref{eq:infinityBC} and \ref{eq:interfaceKinematicsCondition}),
\begin{equation}
    \int_{\xi_0}^\infty \{|\tilde{v}'|^2 + [1 + \tilde{u}''/\tilde{u}]|\tilde{v}|^2 \} \dif \xi = [\tilde{v}^* \tilde{v}']_{\xi_0}^\infty = -\tilde{u}_0\tilde{v}_0'.
\end{equation}
Hence, the imaginary part of eq.~\eqref{eq:alphaBetaCondition} is
\begin{equation}
    \beta = \Im(\tilde{u}_0\tilde{v}_0') = -\Im\left( \int_{\xi_0}^\infty(\tilde{u}''/\tilde{u})|\tilde{v}|^2\dif \xi \right).
    \label{eq:firstExpressionBeta}
\end{equation}
By calculus of residues \cite{miles57}, this expression can be evaluated to
\begin{equation}
    \beta = -\pi |\tilde{v}_c|^2 \frac{\tilde{u}_c''}{\tilde{u}_c'} = -\pi\left|\frac{v_c}{k_x\eta U_1}\right|^2\frac{1}{k}\frac{U_c''}{U_c'},
\end{equation}
where the subscript $c$ denotes values at the location where $\tilde{u} = 0$, $\xi = \xi_c$. To find an expression for $\tilde{v}_c$, eq.~\eqref{eq:uvODE} can be re-written as 
\begin{equation}
    (\tilde{u}\tilde{v}' - \tilde{u}'\tilde{v})' = \tilde{u} \tilde{v}.
\end{equation}
Integrating between $\xi = \xi_c$ and $\xi = \infty$ gives,
\begin{equation}
    \tilde{v}_c = \frac{1}{\tilde{u}_c'}\int_{\xi_c}^\infty \tilde{u} \tilde{v} \dif \xi.
\end{equation}

We now use the same approximate solution of $\tilde{v}$ as above. This approximation gives  
\begin{equation}
    \beta = -\pi\frac{\tilde{u}_c''}{\tilde{u}_c'^3}\left[\int_{\xi_c}^\infty e^{-\xi} \tilde{u}^2 \dif \xi\right]^2.
    \label{eq:general_beta}
\end{equation}
The solution to the full equation has been given by Conte \& Miles \cite{conte59}. Assuming a logarithmic profile, $U = u_\tau/\kappa \log(y/z_0) \implies \tilde{u} = \log(\xi/\xi_c)$. We have here set the reference velocity, $U_1 = u_\tau/\kappa$, where $\kappa$ is the von Kármán constant. With eq.~\eqref{eq:general_beta},
\begin{equation}
    \beta = \pi\xi_c\left[\int_{\xi_c}^\infty e^{-\xi} \log(\xi/\xi_c)^2 \dif \xi\right]^2.
    \label{eq:beta}
\end{equation}

\subsection{Condition for wave growth}
\begin{figure}[H]
  \centering
  \begin{subfigure}{0.45\textwidth}
    \includegraphics[width=7cm]{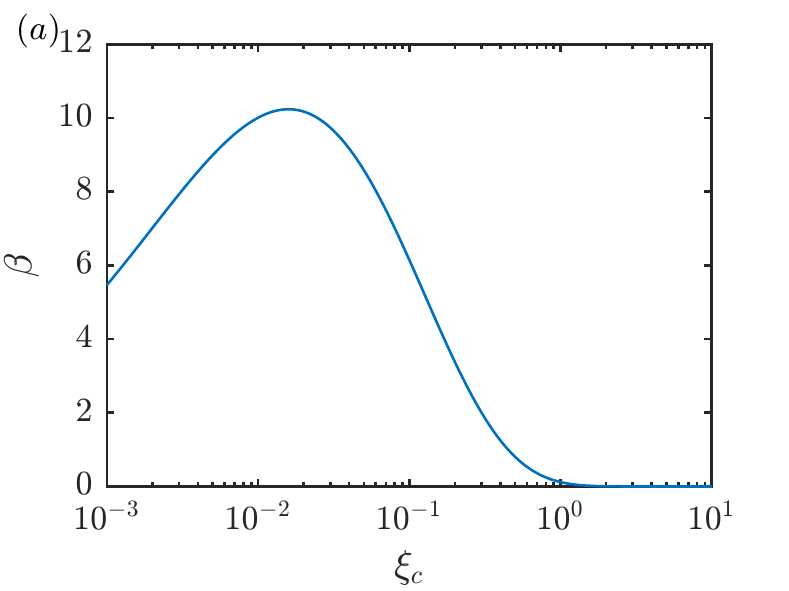}
    \captionlistentry{}
  \end{subfigure}
  \begin{subfigure}{0.45\textwidth}
    \includegraphics[width=7cm]{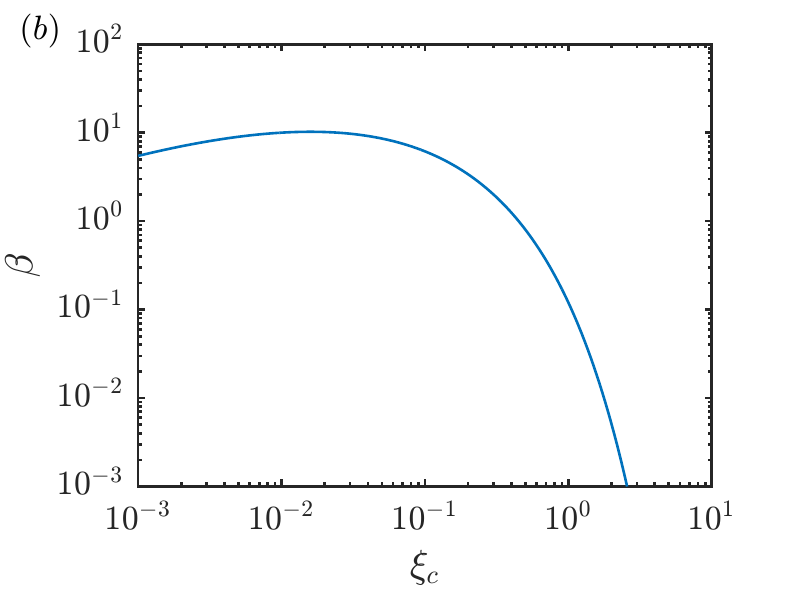}
    \captionlistentry{}
  \end{subfigure}
  \caption{Plot of $\beta$ verses $\xi_c = ky_c$, with ($a$) linear and ($b$) logarithmic vertical axis.}
  \label{fig:beta}
\end{figure}

To the first order, the imaginary wave speed corresponds to 
\begin{equation}
    c_i = c_w\frac{1}{4}\beta\frac{U_1^2}{c_w^2},
\end{equation}
giving wave growth when non-zero. The relation for $\beta$, eq.~\eqref{eq:beta}, can be computed and is plotted in fig.~\ref{fig:beta}. There is a sharp decrease in $\beta$ at about $\xi_c \approx 1$. This means that in order to have amplitude growth,
\begin{equation}
    \xi_c = ky_c \lesssim 1 \implies y_c \lesssim \frac{1}{k} < \frac{w}{\pi},
\end{equation}
where we use the upper bound of $1/k$ as given by eq.~\eqref{eq:wavenumberLimit}. 

\section{Space-time correlations}
\begin{figure}[H]
    \centering
    \begin{subfigure}{0.40\textwidth}
      \centering
      \includegraphics[width=5.5cm]{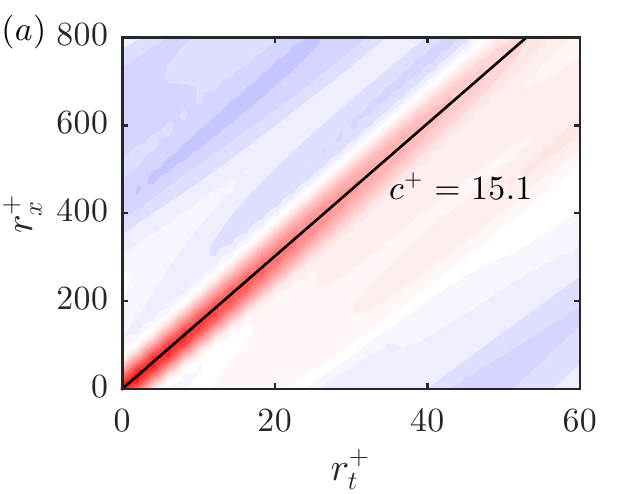}
       \captionlistentry{}
      \label{fig:spaceTimeWe100}
    \end{subfigure}
    \begin{subfigure}{0.4\textwidth}
      \centering
      \includegraphics[width=5.5cm]{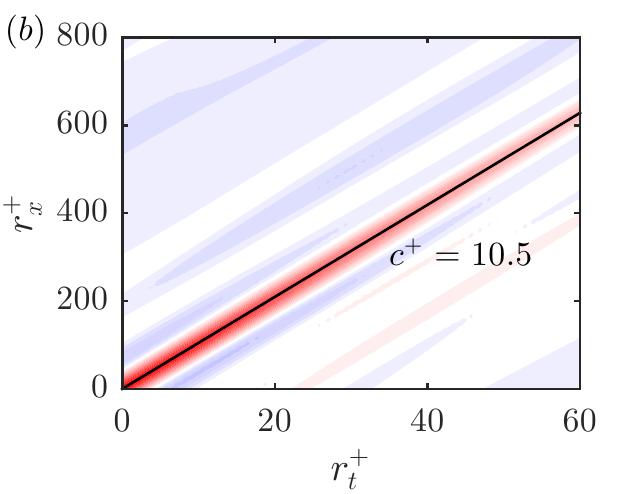}
      \captionlistentry{}
      \label{fig:spaceTimeWe200}
    \end{subfigure}
    \begin{subfigure}{0.1\textwidth}
      \centering
      \vspace{-0.3cm}
      \includegraphics[width=1.6cm]{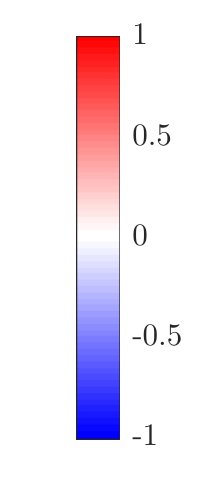}
      \captionlistentry{}
      \label{fig:spaceTimeColorbar}
    \end{subfigure}
    \caption{Space-time correlation of interface height in the middle section of a groove, for $\mu_i/\mu_\infty = 1$ and ($a$) $\WeNumber = 100$ and ($b$) $\WeNumber = 200$. Streamwise displacement is denoted $r_x^+$ and time difference $r_t^+$. The slope of the region with highest correlation corresponds to the dominating phase speed, marked by a black line.}
    \label{fig:spaceTime}
\end{figure}

Space-time correlations are shown in fig.~\ref{fig:spaceTime} for $\mu_i/\mu_\infty = 1$ and $\WeNumber = 100$ and $200$.

\bibliographystyle{ieeetr} \bibliography{references}